\documentclass[useAMS,usegraphicx]{mn2e}
\usepackage{soul}

\usepackage{aas_macros}
\usepackage{rotating}

\title[Multi-wavelength identifications for HerMES]{The {\it Herschel} Multi-Tiered Extragalactic Survey: Source Extraction and Cross-Identifications in Confusion-Dominated SPIRE Images }
\author[I.G.~Roseboom et al.]
{\parbox{\textwidth}{I.G.~Roseboom,$^{1}$\thanks{E-mail: \texttt{i.g.roseboom@sussex.ac.uk}}
S.J.~Oliver,$^{1}$
M.~Kunz,$^{2}$
B.~Altieri,$^{3}$
A.~Amblard,$^{4}$
V.~Arumugam,$^{5}$
R.~Auld,$^{6}$
H.~Aussel,$^{7}$
T.~Babbedge,$^{8}$
M.~B{\'e}thermin,$^{9}$
A.~Blain,$^{10}$
J.~Bock,$^{10,11}$
A.~Boselli,$^{12}$
D.~Brisbin,$^{13}$
V.~Buat,$^{12}$
D.~Burgarella,$^{12}$
N.~Castro-Rodr{\'\i}guez,$^{14,15}$
A.~Cava,$^{14,15}$
P.~Chanial,$^{8}$
E.~Chapin,$^{16}$
D.L.~Clements,$^{8}$
A.~Conley,$^{17}$
L.~Conversi,$^{3}$
A.~Cooray,$^{4,10}$
C.D.~Dowell,$^{10,11}$
E.~Dwek,$^{18}$
S.~Dye,$^{6}$
S.~Eales,$^{6}$
D.~Elbaz,$^{7}$
D.~Farrah,$^{1}$
M.~Fox,$^{8}$
A.~Franceschini,$^{19}$
W.~Gear,$^{6}$
J.~Glenn,$^{17}$
E.A.~Gonz\'alez~Solares,$^{20}$
M.~Griffin,$^{6}$
M.~Halpern,$^{16}$
M.~Harwit,$^{21}$
E.~Hatziminaoglou,$^{22}$
J.~Huang,$^{23}$
E.~Ibar,$^{24}$
K.~Isaak,$^{6}$
R.J.~Ivison,$^{24,5}$
G.~Lagache,$^{9}$
L.~Levenson,$^{10,11}$
N.~Lu,$^{10,25}$
S.~Madden,$^{7}$
B.~Maffei,$^{26}$
G.~Mainetti,$^{19}$
L.~Marchetti,$^{19}$
G.~Marsden,$^{16}$
A.M.J.~Mortier,$^{8}$
H.T.~Nguyen,$^{11,10}$
B.~O'Halloran,$^{8}$
A.~Omont,$^{27}$
M.J.~Page,$^{28}$
P.~Panuzzo,$^{7}$
A.~Papageorgiou,$^{6}$
H.~Patel,$^{8}$
C.P.~Pearson,$^{29,30}$
I.~P{\'e}rez-Fournon,$^{14,15}$
M.~Pohlen,$^{6}$
J.I.~Rawlings,$^{28}$
G.~Raymond,$^{6}$
D.~Rigopoulou,$^{29,31}$
D.~Rizzo,$^{8}$
M.~Rowan-Robinson,$^{8}$
M.~S\'anchez Portal,$^{3}$
B.~Schulz,$^{10,25}$
Douglas~Scott,$^{16}$
N.~Seymour,$^{28}$
D.L.~Shupe,$^{10,25}$
A.J.~Smith,$^{1}$
J.A.~Stevens,$^{32}$
M.~Symeonidis,$^{28}$
M.~Trichas,$^{23}$
K.E.~Tugwell,$^{28}$
M.~Vaccari,$^{19}$
I.~Valtchanov,$^{3}$
J.D.~Vieira,$^{10}$
L.~Vigroux,$^{27}$
L.~Wang,$^{1}$
R.~Ward,$^{1}$
G.~Wright,$^{24}$
C.K.~Xu$^{10,25}$ and
M.~Zemcov$^{10,11}$}\vspace{0.4cm}\\
\parbox{\textwidth}{$^{1}$Astronomy Centre, Dept. of Physics \& Astronomy, University of Sussex, Brighton BN1 9QH, UK\\
$^{2}$D\'epartement de Physique Th\'eorique, Universit\'e de Gen\`eve, 1211 Geneva 4, Switzerland\\
$^{3}$Herschel Science Centre, European Space Astronomy Centre, Villanueva de la Ca\~nada, 28691 Madrid, Spain\\
$^{4}$Dept. of Physics \& Astronomy, University of California, Irvine, CA 92697, USA\\
$^{5}$Institute for Astronomy, University of Edinburgh, Royal Observatory, Blackford Hill, Edinburgh EH9 3HJ, UK\\
$^{6}$Cardiff School of Physics and Astronomy, Cardiff University, Queens Buildings, The Parade, Cardiff CF24 3AA, UK\\
$^{7}$Laboratoire AIM-Paris-Saclay, CEA/DSM/Irfu - CNRS - Universit\'e Paris Diderot, CE-Saclay, pt courrier 131, F-91191 Gif-sur-Yvette, France\\
$^{8}$Astrophysics Group, Imperial College London, Blackett Laboratory, Prince Consort Road, London SW7 2AZ, UK\\
$^{9}$Institut d'Astrophysique Spatiale (IAS), b\^atiment 121, Universit\'e Paris-Sud 11 and CNRS (UMR 8617), 91405 Orsay, France\\
$^{10}$California Institute of Technology, 1200 E. California Blvd., Pasadena, CA 91125, USA\\
$^{11}$Jet Propulsion Laboratory, 4800 Oak Grove Drive, Pasadena, CA 91109, USA\\
$^{12}$Laboratoire d'Astrophysique de Marseille, OAMP, Universit\'e Aix-marseille, CNRS, 38 rue Fr\'ed\'eric Joliot-Curie, 13388 Marseille cedex 13, France\\
$^{13}$Space Science Building, Cornell University, Ithaca, NY, 14853-6801, USA\\
$^{14}$Instituto de Astrof{\'\i}sica de Canarias (IAC), E-38200 La Laguna, Tenerife, Spain\\
$^{15}$Departamento de Astrof{\'\i}sica, Universidad de La Laguna (ULL), E-38205 La Laguna, Tenerife, Spain\\
$^{16}$Department of Physics \& Astronomy, University of British Columbia, 6224 Agricultural Road, Vancouver, BC V6T~1Z1, Canada\\
$^{17}$Dept. of Astrophysical and Planetary Sciences, CASA 389-UCB, University of Colorado, Boulder, CO 80309, USA\\
$^{18}$Observational  Cosmology Lab, Code 665, NASA Goddard Space Flight  Center, Greenbelt, MD 20771, USA\\
$^{19}$Dipartimento di Astronomia, Universit\`{a} di Padova, vicolo Osservatorio, 3, 35122 Padova, Italy\\
$^{20}$Institute of Astronomy, University of Cambridge, Madingley Road, Cambridge CB3 0HA, UK\\
$^{21}$511 H street, SW, Washington, DC 20024-2725, USA\\
$^{22}$ESO, Karl-Schwarzschild-Str. 2, 85748 Garching bei M\"unchen, Germany\\
$^{23}$Harvard-Smithsonian Center for Astrophysics, MS65, 60 Garden Street,  Cambridge,  MA02138, USA\\
$^{24}$UK Astronomy Technology Centre, Royal Observatory, Blackford Hill, Edinburgh EH9 3HJ, UK\\
$^{25}$Infrared Processing and Analysis Center, MS 100-22, California Institute of Technology, JPL, Pasadena, CA 91125, USA\\
$^{26}$School of Physics and Astronomy, The University of Manchester, Alan Turing Building, Oxford Road, Manchester M13 9PL, UK\\
$^{27}$Institut d'Astrophysique de Paris, UMR 7095, CNRS, UPMC Univ. Paris 06, 98bis boulevard Arago, F-75014 Paris, France\\
$^{28}$Mullard Space Science Laboratory, University College London, Holmbury St. Mary, Dorking, Surrey RH5 6NT, UK\\
$^{29}$Space Science \& Technology Department, Rutherford Appleton Laboratory, Chilton, Didcot, Oxfordshire OX11 0QX, UK\\
$^{30}$Institute for Space Imaging Science, University of Lethbridge, Lethbridge, Alberta, T1K 3M4, Canada\\
$^{31}$Astrophysics, Oxford University, Keble Road, Oxford OX1 3RH, UK\\
$^{32}$Centre for Astrophysics Research, University of Hertfordshire, College Lane, Hatfield, Hertfordshire AL10 9AB, UK}}

\begin{document}

\date{\today}

\pagerange{\pageref{firstpage}--\pageref{lastpage}} \pubyear{2010}

\maketitle

\label{firstpage}

\begin{abstract}
We present the cross-identification and source photometry techniques used to process {\it Herschel} SPIRE imaging taken as part of the {\it Herschel} Multi-Tiered Extragalactic Survey (HerMES). Cross-identifications are performed in map-space so as to minimise source blending effects. We make use of a combination of linear inversion and model selection techniques to produce reliable cross-identification catalogues based on {\it Spitzer} MIPS $24\,\mu$m source positions. Testing on simulations and real {\it Herschel} observations show that this approach gives robust results for even the faintest sources ($S_{250}\sim 10$ mJy). We apply our new technique to HerMES SPIRE observations taken as part of the science demostration phase of {\it Herschel}. For our real SPIRE observations we show that, for bright unconfused sources, our flux density estimates are in good agreement with those produced via more traditional point source detection methods (SussExtractor; Savage \& Oliver et al. 2006) by Smith et al 2010. When compared to the measured number density of sources in the SPIRE bands, we show that our method allows the recovery of a larger fraction of faint sources than these traditional methods. However this completeness is heavily dependant on the relative depth of the existing 24$\,\mu$m catalogues and SPIRE imaging. Using our deepest multi-wavelength dataset in GOODS-N, we estimate that the use of shallow 24$\,\mu$m in our other fields introduces an incompleteness at faint levels of between 20--40 per cent at 250 $\mu$m.
\end{abstract}


\section{Introduction}
 Measuring accurate flux densities for sources in astronomical images dominated by confusion noise is the greatest obstacle to scientific analysis of data from next generation telescopes at far-IR to radio wavelengths. Great advances in the sensitivity of instruments at these long wavelengths has meant that the blended signal from numerous, unresolved, faint sources now form a non-negligible fraction of the observed telescope background. Hence confusion noise, i.e. fluctuations in this background, is now the dominant source of noise in deep imaging.

This results in several complications in the analysis of low resolution, long wavelength, imaging. Firstly, confusion acts to increase the positional uncertainty of sources dramatically (e.g. Hogg 2001), making cross-identifications with other wavelengths problematic. Secondly, correlations between the confusing background and sources above the confusion limit result in, at best, flux boosting of detected sources above the confusion limit and, at worst, complex blends of correlated confusion noise, resulting in spurious sources (Scheuer \& Ryle 1957, Condon 1974).

In recent history there have been two distinct approaches to dealing with these issues. Fairly traditional source detection methods, combined with probabilistic approaches for flux boosting and source identification have been used to good effect on sub-mm surveys performed with SCUBA (i.e. Lilly et al. 1999, Mortier et al. 2005, Pope et al. 2005., Ivison et al. 2007).

By comparison others have opted for a more statistical approach, choosing to ignore individual sources and look at the aggregate properties of sources via either stacking (Dole et al. 2006, Pascale et al. 2009, Marsden et al. 2010) or the map statistics themselves via the pixel intensity distribution, the so-called $P(D)$ (e.g. Patanchon et al. 2009). 

Both approaches have advantages and disadvantages; working with individual sources allows the true variation of sub-mm galaxy properties and their correlations with other observables to be properly investigated. However, finding multi-wavelength identifications for individual sub-mm sources is usually difficult, and generally reliable identifications can be found for only a fraction of sources (Ivison et al. 2007; Roseboom et al. 2009). Statistical approaches have the advantage of using all the available data, and hence provide greater precision in the parameters of interest. However, interpretation of these statistically-derived quantities is sometimes complicated, and highly dependent on the choice of parameterisation.%, whether it be the input list in the case of stacking, or the population model, in the case of $P(D)$.

 Recently several authors have made use of an approach which arguably takes the best elements of the three techniques discussed above. By using a linear inversion technique to fit for the flux density of all known sources simultaneously, the ability to work on individual sources is retained, while the information in the map itself can be used to distinguish the contributions from each source. This approach has been used by Scott et al. (2002) to fit the flux densities of SCUBA 850 $\mu$m sources in the 8 mJy survey;  Magnelli et al. (2009) to fit the {\it Spitzer} $24\,\mu$m flux density of IRAC detected sources in GOODS-N; and also by Bethermin et al. (2010) and Chapin et al. (2010) to fit the BLAST and BLAST/LABOCA data, respectively, for $24\,\mu$m detected sources in the extended {\it Chandra} Deep Field South (eCDFS). The key to this approach is its simplicity; the only assumptions are that all sources are unresolved by the telescope, and that the positions of all sources are known. If these assumptions hold, then in the limit of infinite signal-to-noise ratio in the image the resulting flux density measurements would be perfect, irrespective of source density.% Thus this methodology offers the best route to obtaining reliable source flux densities in imaging that is heavily affected by source confusion.

Here we present a similar technique developed to fit the SPIRE band flux densities of $24\,\mu$m sources in fields observed as part of the Science Demonstration Phase (SDP) of the {\it Herschel}\footnote{Herschel is an ESA space observatory with science instruments provided by Principal Investigator consortia. It is open for proposals for observing time from the worldwide astronomical community.} (Pilbratt et al. 2010) mission by the HerMES. The SPIRE instrument, its in-orbit performance, and its scientific capabilities are described by Griffin et al. (2010), and the SPIRE astronomical calibration methods and accuracy are outlined in Swinyard et al. (2010). Our technique is distinct from those discussed above as we include an additional model-selection stage to ensure that only input sources which are justified by the SPIRE data are retained. This stage helps to alleviate the problem of overfitting, i.e. fitting more sources than there are independant data points to constrain.

%Section \ref{sec:definitions} introduces the definitions and syntax for this work.
Section \ref{sec:data} describes the datasets used in this work, Section \ref{sec:linear} presents the linear model used to describe the map. Section \ref{sec:selection} discusses how model selection can be used to ``tune'' the input list of positions, while Sections \ref{sec:simresults} and \ref{sec:xidvsscat} present and discuss the results obtained by implementing this technique on both simulated and observed {\it Herschel} datasets.

\section{Data}\label{sec:data}
In this paper we make use of {\it Herschel} data from HerMES taken as part of the SDP of the {\it Herschel} mission. HerMES performed observations of 5 fields during SDP; these observations are described in Oliver et al. (2010a, 2010b) and summarised in Table \ref{tab:sdpObs}.
\begin{table*}
\caption{HerMES SDP Observations (Oliver et al. 2010a, 2010b). Size is approximate extent of region with uniform coverage. Repeats is total number of pairs of scans in both A and B directions. Sensitivity is that for a point source, ignoring confusion noise. $S_{24}$ refers to existing MIPS observations in these fields.}             % title of Table
\label{tab:sdpObs}      % is used to refer this table in the text
\centering                          % used for centering table
\begin{tabular}{l c rrrrrr}        % centered columns (4 columns)
\hline\hline                 % inserts double horizontal lines
Field Name  &  Size &     RA  & Dec & Repeats &  $S_{250}$ & $S_{24}$\\ 
  &   &  deg & deg & &  (mJy, 5$\sigma$) &  ($\mu$Jy, 95 per cent completeness)\\
\hline                    

A2218    &  $9^\prime\times9^\prime$ &  248.98   &    66.22      &  2 & 2.5 & N/A\\
GOODS-N     &   $30^\prime\times30^\prime $  &189.23   &    62.24 &   30 & 4. & 50\\
LH-North  &    $ 35^\prime\times35^\prime$   &    161.50 &      59.02    &   7 &  8. & 80\\
FLS      &  $155^\prime\times135^\prime$  &  258.97 &      59.39  &     2 & 12.5 & 400\\
LH-SWIRE & $218^\prime \times 218^\prime$ & 162.00   &    58.11  &    2 &  23. & 200\\

\hline                                   %inserts single line
\end{tabular}

\end{table*}

SPIRE data are processed using the {\it Herschel} Interactive Processing Environment ({\tt HIPE}). Details of the SPIRE data processing are described in Smith et al. (2010, in prep.), however we briefly sumarise the main points here. SPIRE maps used in this paper make use of the na\"{i}ve map-making algorithm, with no Wiener filtering applied. While the absolute astrometry of SPIRE imaging is accurate to $\sim2$ arcsec we apply global corrections to the astrometry of the processed maps, based on stacking at the positions of known radio sources. After these corrections have been applied we expect our maps to have an overall astrometric accuracy of $<0.5$ arcsec.

In addition source catalogues are produced using the Sussextractor algorithm in {\tt HIPE} (Savage \& Oliver 2006). Although we do not make use of these catalogues in the cross identification process, comparisons to them are made in Section \ref{sec:xidvsscat}. Throughout we refer to the HIPE processed data products by the moniker SCAT (SPIRE Catalogue) of which we use the latest v3 internal release.

Cross identifications are made between these data, and archival {\it Spitzer} IRAC and MIPS datasets. In the wide, shallow fields, Lockman SWIRE (LH-SWIRE) and FLS, we make use of the multi-wavelength catalogues described in Vaccari et al. (2010, in prep). In Lockman, these catalogues use the {\it Spitzer} SWIRE (Lonsdale et al. 2003) dataset as a starting point, specifically those sources detected by IRAC (at 3.6$\,\mu$m and/or 4.5$\,\mu$m. Analogous catalogues are constructed in FLS, using the IRAC source catalogues of Lacy et al. (2005) and MIPS source catalogues of Fadda et al. (2006). %Counterparts are then identified in SDSS,{\it GALEX}, 2MASS and other public archival optical and near-IR imaging datasets, as well as spectroscopic redshifts from NED\footnote{\url{http://nedwww.ipac.caltech.edu/}}.

In the deeper fields, archival ancillary data are provided by several previous projects. In GOODS-N we make use of the {\it Spitzer} IRAC and MIPS observations taken as part of the GOODS program, specifically the catalogue described in Magnelli et al. (2009), which measures the $24\,\mu$m flux density using the position of IRAC sources as a prior.

In the Lockman North (LH-North) region we use reprocessed archival $24\,\mu$m data from the GO program of Owen et al. % as well as a the compilation of optical/near-IR imaging, photometric and spectroscopic redshifts provided by F.Owen and collaborators. 

\section{Linear Fitting Method} 
\subsection{Basic Equations}\label{sec:linear}
Our data ${\bf d}$ is an image of dimensions $n_1\times n_2=M$ pixels. The pixels are located at discrete positions $({\bf x,y})$.  Our model assumes this data to be formed by a number of point sources with known image coordinates,  ${\bf (u,v)}$, and with unknown flux density, ${\bf f}$.  If each source $i$ makes a contribution to the data given by the point response function (PRF) $P({\bf x}-u_i, {\bf y}-v_i)$ we can describe the flux density in a given pixel $j$ as

\begin{equation}
 d_j=\sum_i{P(x_j-u_i, y_j-v_i)}f_i+\delta_j \label{eqn:lin1}
 \end{equation}
where $\delta_j$ is an additional noise contribution. Thus the entire image {\bf $d$} can be described as:

\begin{equation}
{\bf d}=P({\bf \Delta X},{\bf \Delta Y}){\bf f}+{\bf \delta} ,
\label{eqn:linprf}\end{equation}
where ${\bf {\Delta X}}$ and ${\bf \Delta Y}$ define the offset between pixels and sources.  This is a linear equation of the form
\begin{equation}
  {\bf d}={\bf A}{\bf f}+{\bf \delta}\label{eqn:matrix}.
\end{equation}

Naturally our measures of the pixel intensities {\bf $d$} will have an associated, and measurable, variance and possibly covariance between the pixels, which we define here as ${\bf N_d}=\langle \delta\delta^{\rm T}\rangle$.

To derive the maximum likelihood solution, we write down the
likelihood as the Gaussian probability function for the data given the flux densities:
\begin{eqnarray}
\mathcal{L}(\hat{\bf f}) & =  & p({\bf d}|\hat{\bf f}) \nonumber \\
 &\propto &|{\bf N_d}|^{-1/2} \exp\left\{-\frac{1}{2}({\bf d}-\hat{\bf d})^{\rm T} {\bf N_d}^{-1} ({\bf d}-\hat{\bf d}) \right\} \nonumber 
\end{eqnarray}

\noindent where we define $\hat{\bf d}$ as the data resulting from a given set of flux densities $\hat{\bf f}$. Defining $\chi^2=({\bf d}-\hat{\bf d})^{\rm T} {\bf N}^{-1} ({\bf d}-\hat{\bf d})$ we see that at the maximum of the likelihood we require $\chi^2$ to be at a minimum. However it can be seen that $\hat{\bf d}={\bf A\hat{f}}$, so
\[ \chi^2=({\bf d}-{\bf A}\hat{\bf f})^{\rm T} {\bf N_d}^{-1} ({\bf d}-{\bf A}\hat{\bf f}).\]
Hence at the minimum
\begin{eqnarray}
0 & = & \frac{\partial \chi^2}{\partial {\bf\hat{f}}} \nonumber\\
% & = & \frac{\partial}{\partial {\bf\hat{f}}}\left\{({\bf d}-{\bf A}\hat{\bf f})^{\rm T} {\bf N}^{-1} ({\bf d}-{\bf A}\hat{\bf f})\right\} \nonumber\\
 & = & {\bf A}^{\rm T}{\bf N_d}^{-1}{\bf A}{\bf\hat{f}}-{\bf A}^{\rm T}{\bf N_d}^{-1}{\bf d}, \nonumber
\end{eqnarray}
so the maximum likelihood solution can be written as
\begin{equation}
{\bf \hat{f}}=({\bf A}^{\rm T}{\bf N_d}^{-1}{\bf A})^{-1}\, {\bf A}^{\rm T}{\bf N_d}^{-1}{\bf d},
\label{eqn:mlsolution}
\end{equation}

An equation which is familiar from maximum likelihood map-making for both sub-mm and CMB experiments (e.g. Tegmark 1997, Patanchon et al. 2008, Cantalupo et al. 2010).

%Where ${\bf N}$ is the covariance matrix. 
%Written explicitely in terms of components (with summation over repeated indices
%implied), this becomes
%\[
%\frac{\partial}{\partial f_m} \left[ (d_i-A_{ij} f_j) (N^{-1})_{ik} (d_k-A_{kl} f_l) \right]
%= -2 A_{im} (N^{-1})_{ik}(d_k-A_{kl} f_l) = 0
%\]
%The solution of this equation is the one given above.

\subsection{Estimating the errors}
As this is a linear system the Fisher information matrix can be seen to be 
\begin{equation}
\label{eqn:fim}
{\bf \mathcal{I}} = {\bf A}^{\rm T}{\bf N_d}^{-1}{\bf A} \leq {\bf N_f}^{-1},
\end{equation}

\noindent which, by the Cram\'{e}r-Rao inequality, is the inverse of the lower limit of the covariance matrix of the source flux densities (${\bf N_f}$). Thus the covariance matrix is simply the inverse of the matrix {\bf A$^{\rm T}$N$_d^{-1}$A} in Equation \ref{eqn:mlsolution}. Intuitively this makes sense, if there are no overlaps between the sources then ${\bf \mathcal{I}}$ would be a diagonal matrix with each entry corresponding to {\bf PRF}$^{\rm T}{\bf N_d}^{-1}${\bf PRF} i.e. $\sum PRF_i^2/\langle \delta_i^2\rangle$, where $\langle\delta_i\rangle$ is again the noise in a pixel $i$.

While we can solve Equation \ref{eqn:mlsolution} for the flux densities {\bf $f$} via some fast iterative method, to get the variances we must invert ${\bf \mathcal{I}}$ by `brute-force'. However, the matrix is positive symmetric and highly optimised inversion codes for this class of matrix exist. Here we invert the ${\bf \mathcal{I}}$ directly using {\tt LAPACK/BLAS} routines. 

One drawback to this approach is that we will always be limited to the lower limit of the covariance matrix, given the inequality presented in Equation \ref{eqn:fim}. One alternative would be to use Monte Carlo Markov Chain (MCMC) methods to fully map the posterior probability distribution, allowing the true variance to be properly characterised, as will be discussed in Section \ref{sec:future} %Implementation of MCMC inversion techniques   %These will be discussed further in Section \ref{sec:simresults}

\subsection{Background or other source terms}
We can consider other additive model contributions to the signal in an obvious way, by including extra terms in Equation \ref{eqn:linprf} and calculating the matrix ${\bf A}$ accordingly, with the vector ${\bf f}$ then representing all the model parameters. For example a constant flat background would be a single extra element in the vector ${\bf f}$ with a corresponding row of $(1,1,1,...,1)$ in the matrix ${\bf A}$. More complicated model backgrounds with more parameters can be included by adding extra terms to ${\bf f}$ and ${\bf A}$. The ability to do this is particularly useful for our application to {\it Herschel} SPIRE imaging, as some astronomical flux is lost when removing the telescope background in the map making process for SPIRE imaging.

\section{Optimising the input list}\label{sec:selection}
The linear technique should return the optimal solution for a complete input list, containing the precise position of every source contributing flux to the map. In practice we can never have a precise input list, because some sources will be missing due to flux density limits or masking in the ancillary data while some sources we include may in fact be spurious or emit no flux at the wavelength under investigation. A further complication is that most sub-mm facilities such as {\it Herschel} are not designed as absolute flux measuring devices; the mean level is lost when removing the telescope background. Hence the zero point of the map does not correspond to zero flux density, but rather an unknown mean level, and the faintest sources will appear as fluctuations about this point.

These issues become problematic at high source density, as degenerate solutions to the linear problem become more common. To highlight this consider the most extreme case, using deep optical catalogues as an input to {\it Herschel} SPIRE imaging. The number density of optical sources with $B<28$ is roughly $10^6$ deg$^{-2}$ (Furusawa et al. 2008); given that the SPIRE beam size at $250\,\mu$m is $\sim 3\times 10^{-5}$ deg$^{2}$ the expected number of optical sources per SPIRE beam is $\sim30$. Of course not all of these are going to be luminous at $250\,\mu$m so we need some way of culling those sources which are too faint to be present in our maps.% to 1--2 per beam.

There are two clear approaches to reducing the input list. One method would be to consider properties of the input list, such as the probability of chance alignment given the number density of sources of that flux density (e.g. Downes et al. 1986, Lilly et al. 1999) or the likelihood that a particular source would be sub-mm luminous given its multi-wavelengths properties (e.g.  Pope et al. 2006, Yun et al. 2008, Roseboom et al. 2009).

An alternative is to let the sub-mm data discriminate. The matrix {\bf A} in Equation \ref{eqn:mlsolution} is essentially a model we are trying to fit to the data, with the number of free parameters equal to the number of input sources. However we need to consider the possibility there may be a better model which needs fewer free parameters (sources) to sufficiently describe the data. The use of model selection techniques such as this is common and often rely on criteria such as the Bayesian information criterion (BIC; Schwartz 1978) or the Akaike information criterion (AIC; Akaike 1974). 

Both approaches have advantages at different angular scales. The first approach, which tries to calculate the probability of a chance superposition, is heavily biased towards bright counterparts, and ignores any possible correlations in the clustering at different wavelengths. However the latter approach of letting the data discriminate will not give good results for heavily blended sources (i.e., source seperation much less than the beam FWHM). Thus we want an approach which will incorporate the best elements of both techniques, as detailed in the following sections.

\subsection{Segmenting the map}\label{sec:modelselection}
The biggest problem with implementing any model selection approach to source detection and extraction is that in a na\"{i}ve implementation the number of calculations required is $2^N$, where $N$ is the number of sources to be considered. Given that a typical 1 deg$^2$ map may contain many thousands of sources, some cost saving measures must be introduced. The first step is to segment the map into isolated regions in which sources may contribute significantly, but not affect other sources outside it. For this step we can re-use the Fisher information matrix calculated in Equation \ref{eqn:mlsolution}, {\bf$ \mathcal{I}=A^{\rm T}N^{-1}A$}. The non-diagonal elements of ${\bf \mathcal{I}}$, $i,j$ for $j\ne i$, describe the fractional contributions the source $j$ makes to the noise-weighted, PRF convolved flux density in the map at the position of source $i$, i.e. $\sum_j \mathcal{I}_{i,j}f_j = m_i$, where $m_i$ is the flux density in a PRF convolved map at the position of source $i$, weighted by the pixel noise. Thus we isolate regions of blended sources by taking sources to be paired if
\[\mathcal{I}_{i,j}m_j>1,\]

\noindent i.e. if the flux density contributed by one source to another is greater than the 1$\sigma$ pixel noise. 

One problem with this approach is that it assumes we already know the flux densities of the sources in the map. However, given that we are only trying to segment the map, it should suffice to use some initial estimate of the flux densities {\bf $f$}. Here we choose to simply use the PRF convolved flux density at the position of source $i$, irrespective of its neighbouring sources, i.e {\bf $f_0=A^{\rm T}N^{-1}d/\mathcal{I}_{\rm diag}$} where only the diagonal elements of {\bf $\mathcal{I}$} are considered. In this framework {\bf $f_0$} can be recognised as the upper limits to the flux densities. Chains of connected sources are identified by starting at one source, and going through all the elements of its row in {\bf$\mathcal{I}$}, grouping the connected sources. After this first step iterations of this same process are repeated on all the sources in the group, until the group does not continue to grow in size.

\subsection{Using model selection for source detection}
Once the map has been segmented into groups we can use model selection to decide which sources in each group are justified by the data. However, the number of calculations to be performed is still $2^{N}$, where $N$ is now the group size. As discussed above, in heavily confused images the number density of input sources could be as high as 10 per beam element, resulting in a very large number of calculations to be performed. As an alternative we adopt an iterative `top-down' approach in which we jackknife the input list i.e. consider all the models which have $N-1$ sources, and select the best model. The process is then repeated with $N-2, N-3,...,N-s$ sources, until a better model cannot be found. Models are compared using the Akaike Information Criteria, corrected for finite sample sizes (AICc);
\[{\rm AICc}=2k-2~\ln(\mathcal{L})+\frac{2k(k+1)}{n-k-1},\]
Where $k$ is the number of parameters, $\mathcal{L}$ is the likelihood, and $n$ is the total number of data points used in the fit. While the BIC could have been used, here we choose the AIC as it penalises extra parameters less harshly than the BIC. Since our parameters are actual known sources (as opposed to simply free parameters in a model) we have good reason to believe they should be included unless there is evidence to the contrary. For our source fitting $-2\ln(\mathcal{L})=\chi^2$ model so the AICc becomes:
\[{\rm AICc}=\chi^2+2k+\frac{2k(k+1)}{n-k-1},\] 

\noindent where $\chi^2$ is calculated on the fit to the map segment. Since we are fitting in source space, $n$ here is the original number of sources considered and in the first step $k=n$. 
\subsection{Weighting the input list}
A number of well-established probabilistic approaches exist for weighting identifications of low resolution, sub-mm sources. We require something more, because the traditional techniques require a source detection stage, which is absent from our methodology. What we wish to know is the likelihood that a particular input source is luminous (or more practically, detectable) in the sub-mm band of interest. One way to do this would be to consider the existing full multi-wavelength dataset for each input source, and predict the sub-mm flux density and its variance from the full range of plausible SEDs. While this would in principle return the best results, implementation of such an approach would be difficult and give mixed success due to the heterogeneous nature of most multi-wavelength data sets. 

A simpler alternative is to weight the models by how likely they would be to appear by chance, i.e. what is the likelihood that a source is a random superposition? This approach is analagous to the `$p$ statistic' analysis (Downes et al. 1986); however, in our implementation we do not have positions for our sub-mm sources and hence cannot work out the probability of finding a counterpart within a given search radius and separation. 

Since the AIC offers a relative comparison of models, the absolute likelihood here is not important, thus we introduce a more na\"{i}ve, but useful, estimate of the probability of a chance alignment. For a given source $i$ we calculate the probability $\phi_i$ of finding a source in the input catalogue with flux density $F$ greater than the source under consideration $F_{i}$ within an area of one beam element $A=\pi{\rm FWHM}^2/4\ln2$:
\[\phi_i=M_{F>F_{i}}~A.\]

Where $M_{f>f_{\rm in}}$ is the number density of sources present in the input list with $F>F_{i}$. We add this probability to the model selection stage and hence the AICc calculated for each model becomes;
\[AICc=\chi^2+2k+2\times \ln\left(\sum \phi_{i}\right)+\frac{2k(k+1)}{n-k-1}\] 

\noindent where the $\sum \phi_{i}$ runs over all of the sources which are assigned zero flux density via the model testing or the fitting process itself.

\section{The HerMES Cross-Identification Algorithm}\label{sec:HermesXID}
As HerMES will identify $>200{,}000$ sources in the {\it Herschel} SPIRE bands across all of our survey fields (Oliver et al. 2010a), we need an algorithmic, machine-based, approach to producing cross-identifications (XIDs) across the many data sets present in our fields. To achieve this we utilise an implementation of the method described above. One of the key features of HerMES is that all of the planned survey fields contain existing {\it Spitzer} data from a range of legacy surveys. More importantly the tiered nature of HerMES is well matched to the variable quality of the {\it Spitzer} data, in particular the MIPS $24\,\mu$m observations. This is highlighted by comparing the $S_{250}$ and $S_{24}$ sensitivities in Table \ref{tab:sdpObs}. With the exception of FLS, all of the SDP observations have a limiting $S_{250}/S_{24}$ colour of $\sim 100$. Using a compilation of pre-{\it Herschel} emphirical models (Fernadez-Conde et al. 2008; Le Borgne et al 2009; Franceschini et al., in prep; Pearson et al., in prop; Valiante et al. 2010; Xu et al. 2001) we estimate that 0.4--24 per cent of $S_{250}>1$ mJy sources have $S_{250}/S_{24}>100$, with the majority of these (up to 70 per cent) lying in the range $1.2<z<1.6$ where the $24\,\mu$m band is coincident with the $10\,\mu$m silicate feature present in strong absorption in typical starburst galaxies.

It is also clear from existing measurements of the cosmic IR background (CIRB) from BLAST (Devlin et al. 2009) that sources already detected at 24$\,\mu$m with {\it Spitzer} are the dominant contributor at these wavelengths. In particular, Pascale et al. (2009) show that greater than $90$ per cent of the CIRB at BLAST/SPIRE wavelengths can be accounted for by $24\,\mu$m sources with $S_{24} \ge 100~\mu$Jy. Hence we can be confident that using the $24\,\mu$m source lists as a model for the positions of sources in the SPIRE maps is appropriate. It is also worth considering that in the deepest fields (i.e. GOODS-N) the source density of $24\,\mu$m sources is $\sim24{,}000$ deg$^{-2}$, or $\sim2$ SPIRE $250\,\mu$m beam elements per source. Thus even recovering the SPIRE fluxes for the detected $24\,\mu$m sources, involves going significantly beyond the confusion limit. 

The full algorithm used to produce the XID catalogue for the SDP observations is described below. While it would be possible to include Gaussian priors on the SPIRE flux densities, at this stage we do not understand the SEDs of our SPIRE sources well enough to accurately predict the range of SPIRE flux densities from the existing {\it Spitzer} and short wavelength data (i.e. Rowan-Robinson et al. 2010). Hence only the simple non-negative flux density prior, $S_{\rm SPIRE} \ge 0$, has been implemented. This is achieved by using the Bounded Variable Least Squares (BVLS) algorithm described by Stark \& Parker (1995) to perform the matrix inversion. 

 It should be noted that this algorithm has been developed in parallel with the other data reduction techniques (i.e. Smith et al. 2010; Levenson et al. 2010) for use in the first SDP science papers from HerMES. Thus while this approach has proven to give the best performance under testing, it is clear that several aspects could be easily improved. However, to maintain consistency with the results presented in other HerMES SDP papers we only consider our original algorithm in the following. A description of problematic aspects of this approach, and how they may be improved in future applications, is presented in Section \ref{sec:future}.

%\vspace{0.3cm}
\subsection{Step-by-step description of HerMES XID algoritm}
\noindent For the $250\,\mu$m band we follow these specific steps:
%\vspace{-0.1cm}
\begin{description}

\item[1)] Produce input list from available $24\,\mu$m source catalogues. Sources are considered if they are detected at 5$\sigma$ in the MIPS $24\,\mu$m imaging, and if they are above a given flux density limit: 20 $\mu$Jy for GOODS-N, 50 $\mu$Jy for LH-North, and 200 $\mu$Jy for FLS and LH-SWIRE

\item[2)] Calculate the matrices needed for the inversion method using the input list, PRF model, and SPIRE $250\,\mu$m map and variance (i.e. ${\bf A}$, ${\bf N}$ and ${\bf d}$ from Equation \ref{eqn:mlsolution}).
\item[3)] Generate the matrix ${\bf \mathcal{I}}={\bf A}^{\rm T}{\bf N}^{-1}{\bf A}$ and ${\bf m}={\bf A}^{\rm T}{\bf N}^{-1}{\bf d}$.
\item[4)] Segment the map using information contained in ${\bf \mathcal{I}}$ and ${\bf m}$. Segments are produced by weighing the contribution from source blending against the instrumental noise, as described in Section \ref{sec:modelselection}.  In practice this method produces segments too large to be solved in a reasonable time, thus we add an extra factor, $\Delta_{d}$, to the instrumental noise in quadrature. For the catalogues described here, $\Delta_{d}=1$ mJy in all cases. Thus sources are segmented into groupings where no external sources contribute more than $(\sigma^2_{inst}+\Delta^2_{d})^{1/2}$ to any given source within the segment. In practice this extra term is thought to be less than the errors introduced by the unknown background (2--3 mJy), and the incompleteness of the input list, characterised by the surface brightness of sources undetected at 24$\,\mu$ (2--3 mJy/beam, predicted by FC08 mocks) and hence has a negligible effect on the quality of the output catalogues. A given source is allocated to exactly one segment, such that the algorithm returns a single estimate of the flux density for each input object. Typical segments are 10--50 sources in size for our deepest fields, with a maxmium size of $\sim200$.%This assertion has been proven by testing $n_{d}=0.$ on subregions o
\item[5)] For each segment:
  \begin{description}\vspace{-0.2cm}
  \item[5.1)] Build the smaller ${\bf \mathcal{I}'}$ and ${\bf m'}$ for this segment from ${\bf \mathcal{I}}$ and ${\bf m}$.
  \item[5.2)] Build the noise weighted mini-map of the segment region ${\bf d'}={\bf d}{{\bf N}^{-1}}$.
  \item[5.2)] Add a local flat background under the segment to ${\bf \mathcal{I}'}$ and ${\bf m'}$. The response of the background is taken to be $1/{M_{\rm src}'}$, where ${M_{\rm src}'}$ is the number of sources in the segment. It is necessary to fit this background to recover some of the astronomical flux lost to the telescope background in the map-making process. Note that we do not allow this parameter to be removed by the model selection stage.
  \item[5.3)] Solve ${\bf \mathcal{I}'}{\bf f'}={\bf m'}$ for source flux densities ${\bf f'}$ using BVLS .
  \item[5.4)] Calculate the initial $\chi^2$ and AICc from the solution in 5.3 and mini-map ${\bf d'}$.
  \item[5.5)] Iteratively search for the minimum AICc, starting with $i=1$:
    \begin{description}
      \item[5.5.1)] Fit the segment with all ${M_{\rm src}'-i}$ combinations;
      \item[5.5.2)] Measure $\chi^2$ and AICc values for each combination;
      \item[5.5.3)] From the set of ${M_{\rm src}'-i}$ AICc values, identify the minimum; 
      \item[5.5.4)] If min[AICc(${M_{\rm src}'-i}$)]$<$min[AICc(${M_{\rm src}'-i+1}$)] then remove the source corresponding to that model from consideration, increment $i$, and go to step 5.5.1. Otherwise go to step 5.6.
        \end{description}
\vspace{-0.2cm}
  \item[5.6)] Calculate the lower limit to the covariance matrix for the sources in this segment by directly inverting ${\bf \mathcal{I}'}$ using LAPACK/BLAS routines. ${\bf \mathcal{I}'}$ at this stage contains only the sources which have not been removed by the model selection stage (5.5). 
  \item[5.7)] Use the covariance matrix to find the maximum absolute value of the Pearson correlation coefficient\footnote{Pearson correlation coefficient is: $r=cov_{i,j}/{\sigma_i}{\sigma_j}$} for sources in the segment. This can be used later to identify heavily blended sources which cannot be recovered by this method.
 \end{description}
\item[6)] Write out the measured quantities (flux density, error, background for each segment, $\chi^2$ and correlation).

\end{description}

\subsection{Initial results from processing HerMES observations}
The HerMES XID algorithm has been used to produce catalogues in each of the SDP fields, with the exception of Abell 2218. The typical fraction of sources removed by the model selection stage is 20--40 per cent, although this is strongly dependent on the depth of the $24\,\mu$m input list. Rare cases do occur where all sources are retained or only 1 source is retained in a segment. To illustrate this, in the GOODS-N field 47 per cent of input sources are rejected by the model selection. These sources have a median $24\,\mu$m flux density of 64 $\mu$Jy while only 5 per cent have a $24\,\mu$m flux density greater than 150 $\mu$Jy. By contrast, 21 per cent of the input sources in LH-SWIRE are rejected, with a median $24\,\mu$m flux density of 260 $\mu$Jy and 5 per cent being greater than 700 $\mu$Jy.

In order to achieve consistency between the 3 SPIRE bands we only carry out the model selection stage of the algorithm for the $250\,\mu$m band.  An alternative approach, where all three SPIRE bands are treated independently, was initially considered, but found to give poor results. In particular the increase in beam size from 250 to 350 to $500\,\mu$m results in a decreased ability to deblend at long wavelengths and a preference to retain fewer sources. This naturally leads to inconsistencies between the measurements in the different bands. Thus it was decided to use the $250\,\mu$m results to determine which sources were indeed present at the SPIRE bands. One downfall of this approach is that some faint sources will be missed, where the observed-frame SED peaks longward of $250\,\mu$m. However, it was decided that these $250\,\mu$m faint sources would be too hard to recover reliably at this stage. Again we can use the pre-{\it Herschel} mock catalogues of Fernandez-Conde et al. (2008) to estimate the number of 350 and 500 $\mu$m sources missed by this requirement. Assuming a uniform sensitivity across the SPIRE bands, and the depth of the deepest field considered here (GOODS-N; 4 mJy), the number of sources the additional incompleteness due to requiring a detection at 250 $\mu$m is only 0.5 per cent. However it is clear that this estimate is highly dependant on the range of SEDs used in the Fernandez-Conde et al. models.

Ideally the model selection would be performed over all three bands concurrently, such that evidence in any one band for a particular source would cause it to preferred by the model. However, it was not possible to implement this approach in time for the SDP papers. 

Of course even with these additions we are still limited to those sources which are detected at 24 $\mu$m. An additional step to find entirely new sources in the residual maps, using the AICc to determine their significance, would rectify this. Again it was not possible to implement such a feature in time for the SDP papers.

For the 350 and $500\,\mu$m bands only sources which are found to have $S_{250}>1\sigma$ are considered. The flux densities for the 350 and $500\,\mu$m sources are then measured using steps 1--5.4 and 6 only.

XID catalogues have been produced in this way for the SDP fields described in Table \ref{tab:sdpObs}. As an input to the algorithm we take the $24\,\mu$m source catalogues described in Section \ref{sec:data} and the known PRF. Testing on bright point sources has shown that the PRF can be adequetely described as a 2D Gaussian with FWHM=18.15, 25.15 and 36.3 arcsec, for the 250, 350 and $500\,\mu$m bands, respectively.% The use of a 2-D Gaussian as opposed to the true PRF shape is thought to introduce flux density errors on the $<1$ per cent level.% substantially smaller than the $\sim15$ per cent calibration accuracy (Griffin et al. 2010).

While the input source list is defined by the $24\,\mu$m flux density limits, we use source positions from {\it Spitzer} IRAC 3.6 $\mu$m imaging where there is deep co-incident data and previous associations between the two data sets have been made. This occurs in all of our fields, with the exception of LH-North, and the wider area of HDF-N. The IRAC positional accuracy is typically $\sim 0.2$ arcsec (as opposed to $\sim 1$ arcsec for 24 $\mu$m) and hence using these eliminates any error in the flux density solutions introduced by astrometric errors. 

The resulting HerMES XID catalogues contain the complete input $24\,\mu$m source catalogue, as well as any previously associated data sets at other wavelengths (see Vaccari et al. 2010, in prep.), as well as the best estimate of the SPIRE flux density for each $24\,\mu$m source passing our input selection criteria.

In addition to the flux density and error in each band for each input $24\,\mu$m source, the SPIRE component of the XID catalogues contain a number of extra columns describing diagnostics of the fitting process and local source confusion. These extra measures include:

\begin{itemize}
\item Maximum absolute value of the Pearson correlation coefficient calculated on the covariance matrix of the flux density solution (hereafter refered to as $\rho$).
\item $\chi^2$ of the source solution in the neighbourhood of source (7 pixel radius).
\item The background level estimated in the fitting.
\item The number of sources in the segment containing this source.
\item The ID number of the segment.
\item The PRF-smoothed flux density at the position of this source, ignoring contributions from neighbouring sources and the background.
\item The number of $24\,\mu$m sources within a radius of the FWHM with greater than 50 per cent of the flux density of this source.
\item The `purity' of the SPIRE flux density, based on the ratio of this source's $24\,\mu$m flux density to the $24\,\mu$m flux density smoothed with the SPIRE PRF at this position (see Brisbin et al. 2010).
\end{itemize}

The reason for including these extra columns is to enable samples of varying quality to be extracted from the XID catalogues based on differing scientific requirements. From an early assessment of the XID algorithm performance the recommended quality cuts for typical science applications were:
\begin{itemize}
\item $S_{\lambda}>5 \times \Delta S_{\lambda}$;
\item $\rho<0.8$;
\item $\chi^2<5$.
\end{itemize}

Given the more detailed analysis presented below, these cuts have proven to return very reliable samples, although possibly at the expense of completeness. Hence they represent fairly conservative guidelines for the use of the XID catalogues.

\section{Testing on Simulations}\label{sec:simresults}

To quantify the effectiveness of these new techniques we consider simulated SPIRE images. Here we consider two simulated cases: a `deep' map, where $\sigma_{\rm conf}\gg\sigma_{\rm inst}$; and a `shallow' map where $\sigma_{\rm inst}\ge\sigma_{\rm conf}$. In each case we simulate a 2.2$^{o}$ $\times$2.2$^{o}$ patch of sky in all three SPIRE bands, taking the mock catalogues of Fernandez-Conde et al. (2008, henceforth FC08) as an input. While many mock catalogues exist at these wavelengths, the FC08 mocks were found to give the best match to the observed confusion noise and source colours in real SPIRE data. One key feature of the FC08 mock catalogues is that they incorporate a prescription for the clustering of sources (albeit flux and SED independent). This characteristic is of particular importance as the clustering introduces correlations between the resolved sources and the confusing background of the sort present in the real data. Additionally the FC08 mocks incorporate a semi-realistic range of SED types, and their evolution, based on a combination of detailed modelling of local sources, and constraints placed by pre-{\it Herschel} number counts at {\it Spitzer}, {\it ISO} and SCUBA wavelengths. 

Simulated maps are produced from the positions and flux densities quoted in the mock catalogues by first making noise-free maps in each band, using the known SPIRE PRF parameters. Secondly, Gaussian noise and a flat background are added. To give the best possible correspondence to the real observations, this second step is repeated, varying the Gaussian noise and background, until the best match to the P(D) in the observed SPIRE maps is found. For the deep scenario we match the observations in our GOODS-N observations, while for the shallow simulation we match to observations in LH-SWIRE. Given the confusion noise at SPIRE bands is known to be $\sim$5--7 mJy (Nguyen et al. 2010, Smith et al. 2010), these scenarios represent the confusion-noise-dominated, and instrument-noise-dominated cases, respectively. 

\begin{figure}

\includegraphics[scale=0.55]{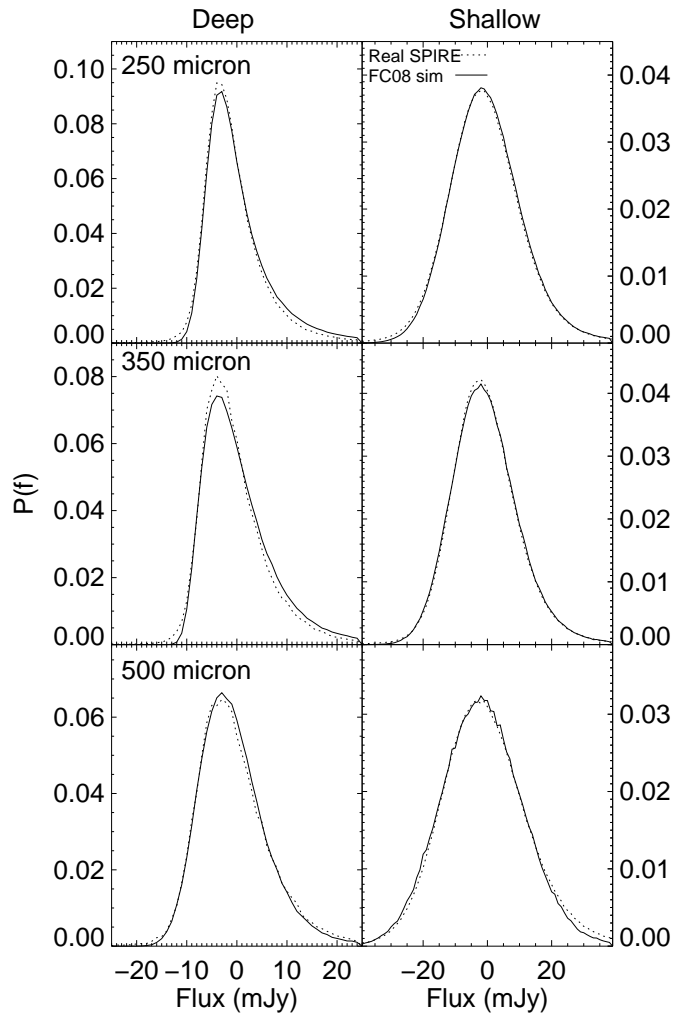}
\caption{Normalised distribution of intensity values in each pixel ({\it P(D)}) for the SPIRE observations in GOODS-N (left; dashed), LH-SWIRE (right;dashed) and our deep and shallow simulated maps (left and right respectively; solid lines), based on Fernandez-Conde et al. (2008). }
\label{fig:simpd}
\end{figure}

Figure \ref{fig:simpd} compares the {\it P(D)} distributions for HerMES SPIRE observations in GOODS-N and LH-SWIRE to corresponding simulations. Table \ref{tab:simnoise} lists the background and Gaussian noise added to each pixel in the simulated map in order to match the observations.
\begin{table}
\caption{Details of simulation parameters}
\label{tab:simnoise}
\begin{tabular}{llllll}
\hline\hline
 & Band & FWHM&  Noise & Background\\
  &  ($\mu$m) & (arcsec) & (mJy beam$^{-1}$) & (mJy beam$^{-1}$)\\
\hline
Deep & 250 & 18.15 &  2. & 6.5\\
Deep & 350 & 25.15 &  1.1 & 10.9\\
Deep & 500 & 36.3 &  2.7 & 14.7\\
Shallow & 250 & 18.15 &  9. & 7.1 \\
Shallow & 350 & 25.15 & 7.5 & 11.9\\
Shallow & 500 & 36.3 & 11.1 & 15.7\\
\end{tabular}
\end{table}

A mock $24\,\mu$m input catalogue is produced by cutting the FC08 simulation at a level representative of the quality of the $24\,\mu$m data in our observed fields: $S_{24}>50 \mu$Jy for the deep simulation, and $S_{24}>200 \mu$Jy for the shallow simulation. In addition to these flux density limits some realistic limitations on source confusion in $24\,\mu$m {\it detected} source lists are imposed.  As the beam size of MIPS 24$\,\mu$m imaging is 6 arcsec very few sources appear with seperations of $<\,$3 arcsec, and those which do quite often turn out to be unreliable. Thus mock source pairs which are seperated by less than half of the {\it Spitzer} MIPS $24\,\mu$m beam (3 arcsec) are filtered, with preference given to the brighter source. Additionally, sources within the Airy profile of bright 24$\,\mu$m sources are also removed. The first Airy ring of the MIPS $24\,\mu$m PRF has a peak level of $\sim$10 per cent of the peak of the PRF\footnote{see \url{http://ssc.spitzer.caltech.edu/mips/mipsinstrumenthandbook/}}. Hence we filter pairs of sources on the scale of 8 arcsec, with flux density ratios of greater than 10, again giving preference to the brighter source. 

Simulated XID catalogues are produced using the HerMES XID algorithm outlined above (herefter refered to as Method A). We compare our approach with two previously adopted XID methods for far-IR datasets using the same simulations: A catalogue-space method, using a combination of the Sussextractor algorithm and $p$-statistic matching (Method B), and a variant of existing linear inversion methods, based on that presented in B\'{e}thermin et al. (2010) (Method C). For method C we filter pairs in the $24\,\mu$m input catalogue at seperations of less than $20$ arcsec, giving preference to the brighter source at 24 $\mu$m, as per  B\'{e}thermin et al. (2010). In addition for method C we make use of a conjugate gradient method with no flux density priors to perform the inversion, as opposed to the BVLS method with a non-negative prior described above.

The performance of source extraction and cross-identification methods are typically characterised by two metrics: the completeness, i.e. the fraction of sources recovered at a given flux density; and the reliability or mis-ID rate. While the notion of completeness translates well to the methods presented here, reliability is not an intuitively useful quantity when performing XIDs in the map-space. We know (or assume) that all of our $24\,\mu$m sources are reliable; the aim is solely to accurately measure their flux densities at other wavelengths. Thus the second metric by which we judge our XID methods is flux density accuracy.%, in terms of both sigma-clipped RMS, and catastrophic failure rate.

In constructing the simulated catalogues for all three methods we need to make some XID and flux density quality cuts.

For the method A we select all sources from the output catalogues. Additionally we define a second sample using the $\chi^2<5$ and $\rho<0.8$ selection thresholds described in Section \ref{sec:HermesXID}. To emphasise the effect these additional cuts have on the completeness and flux density accuracy we denote the use of these additional quality cuts Method A'.

For method B we take all sources in the Sussextractor output lists and try to find matches in the mock $24\,\mu$m catalogue within a search radius of 10 arcsec, 14 arcsec and 20 arcsec for the three SPIRE bands, respectively. For all sources within the search radius we calculate the $p$-statistic of the match using the formula of Downes et al. (1986), taking those with $p<0.1$ to be possible counterparts. Cases are excluded where there are multiple counterparts with $p<0.1$ for a single detected source. 

For method C we take all sources in the output catalogue.% with $S_{\lambda}/\Delta S_{\lambda}>5$.

\begin{figure*}
\includegraphics[scale=0.4]{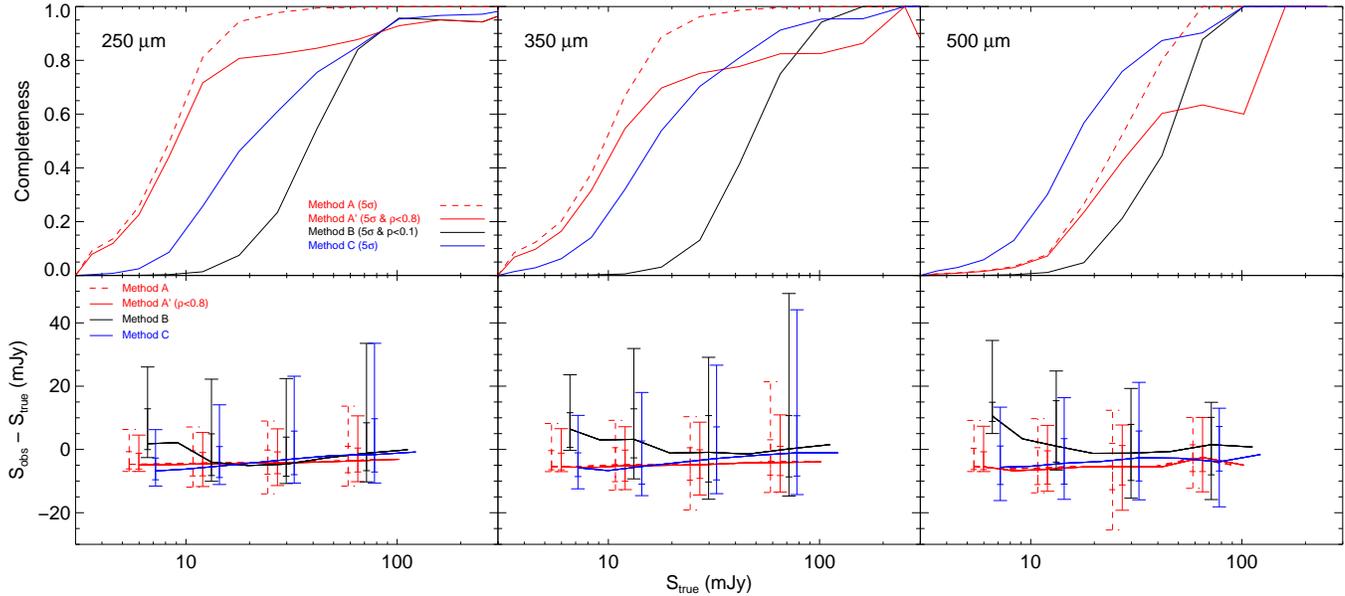}
\caption{Deep simulation results for completeness (top) and flux density error (bottom). Completeness is for 5$\sigma_{\rm catalogue}$ sources only, while flux density error is measured for all objects in the resulting catalogues, irregardless of significance. Results are shown for the three SPIRE bands; $250\,\mu$m (left), $350\,\mu$m (middle) and $500\,\mu$m (right) and the three XID algorithms considered: Method A (HerMES XID algorithm; red dashed line); Method A' (HerMES XID algorithm with $\rho<0.8$ quality cut; red solid line); Method B (Sussextractor+$p$-stat; black); Method C (simple linear inversion method; blue). Lines in bottom panel represent the median flux density error for each band/method, while the error bars are the 1$\sigma$ and $3\sigma$ variation. Both map based, linear inversion methods (methods A and C) are seen to outperform the catalogue based method at faint flux densities. The low completeness of the HerMES XID algorithm at 350 and $500\,\mu$m can be attributed to the $\rho<0.8$ cut.}
\label{fig:simcomp}
\end{figure*}

\begin{figure*}
\includegraphics[scale=0.4]{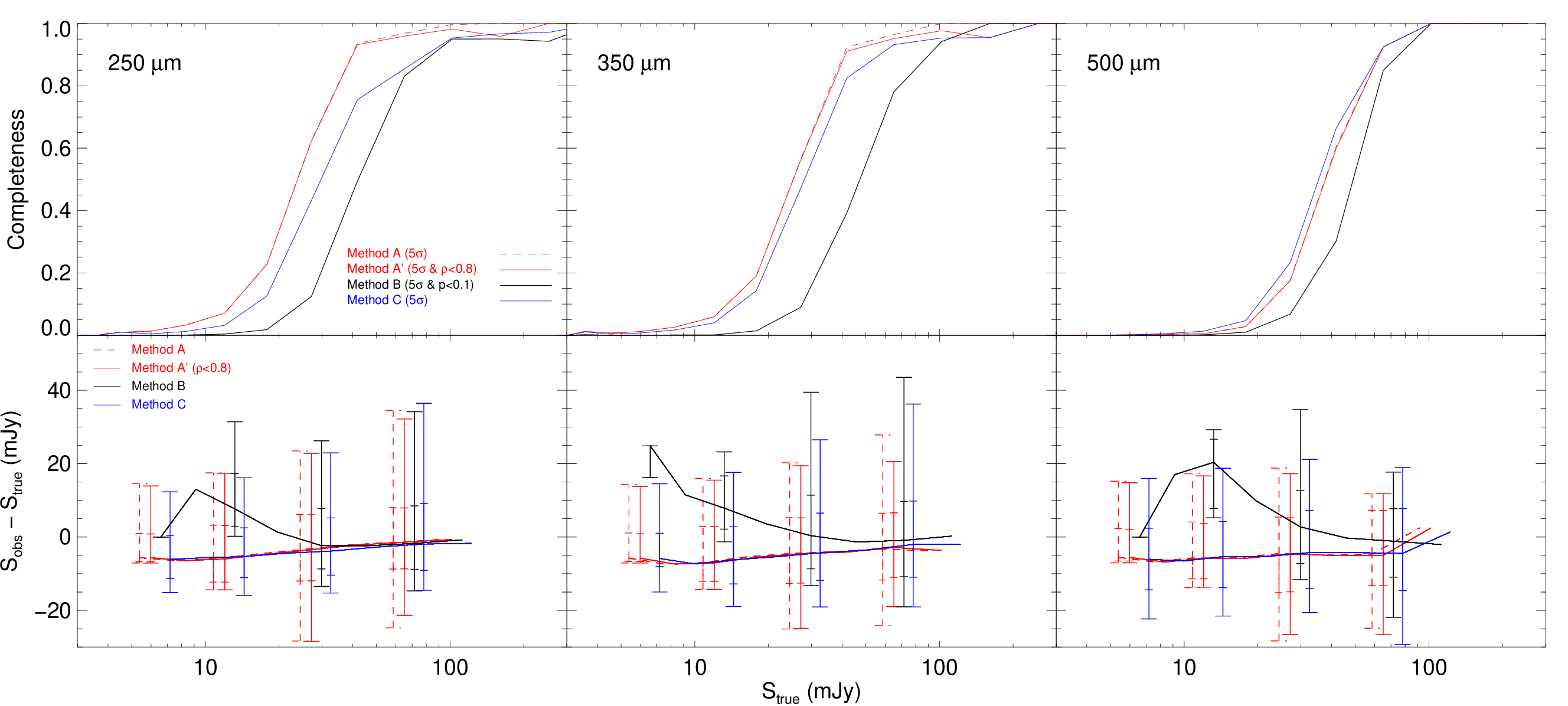}
\caption{Shallow simulation results for completeness (top) and flux density error (bottom).  Completeness is for 5$\sigma_{\rm catalogue}$ sources only, while flux density error is measured for all objects in the resulting catalogues, irregardless of significance. Lines as per Figure \ref{fig:simcomp}. Again both map based, linear inversion methods (methods A and C) are seen to outperform the catalogue based method at faint flux densities. }
\label{fig:simcompshallow}
\end{figure*}

Figures \ref{fig:simcomp} and \ref{fig:simcompshallow} present the completeness and flux density accuracy for the three methods. In both Figures completeness is defined as the fraction of sources in the output catalogues recovered at $5\sigma$ significance and satisfying the above conditions, to the total number of sources in the original FC08 mocks. By contrast the flux density accuracy is measured across all recovered sources, irregardless of significance, although it should be noted that for a source to appear in the Sussextractor list it must be detected by that algorithm at $>3\sigma$. A summary of the key statistics is also presented in Table \ref{tab:simresults}. Both map-based methods (A and C) have higher completeness at faint flux densities at all wavelengths. It should be noted that for the method B it is the requirement that an ID be `secure' that forces the completeness to be low. As shown in Smith et al. (2010), the completeness of the Sussextractor-alone catalogues is comparable to that achieved by the HerMES XID algorithm (method A). In the deep simulations, the $\rho<0.8$ cut imposed in method A' has a similarly dramatic effect on completeness in the 350 and $500\,\mu$m bands. This is primarily due to the lower resolution and $250\,\mu$m-based input list used in the longer wavelength bands. As no model selection stage is performed on the 350 and $500\,\mu$m images we are attempting to fit more sources than can be resolved in the map, leading to strong degeneracies between close pairs. This is understandable, as the typical separation between input sources to the 350 and $500\,\mu$m map is 1.5 and 1 pixel (15 arcsec), respectively. It should be noted that similar degeneracies in the simple linear inversion are removed by the initial spatial filtering of the input list.

The completeness of the method A (and A') is not consistently better than the other methods; which can even be superior at bright flux densities in the long wavelength bands. However, the flux density accuracy of the Hermes XID algorithm is consistently better. This is most striking in the deep simulation, where the flux density accuracy of the method A, and in particular method A', is not only better, but has significantly fewer sources which have been boosted to erroneously high flux densities. 

Another feature which is clear from Figures \ref{fig:simcomp} and \ref{fig:simcompshallow} is that the mean flux density error is always negative for the linear methods, i.e. method A and C systematically underestimate the flux density of sources. The mean S$_{\rm obs}-$S$_{\rm true}$ for each method and simulation is given in Table \ref{tab:simresults}. The origin of these negative offsets can be attributed to fact the maps have a mean of zero, whereas we know that there is an unresolved background of sources contributing to each pixel (or beam) in the maps. For traditional source detection and extraction methods this is preferable, as fluctuations in the confusing background appear as quasi-symmetric noise about zero, and hence can be treated as another pseudo-Gaussian noise term.

However for our Method A and C the number density of our input list is much higher than could not be identified in the map blindly. Hence we are attempting to `resolve' some of the confusing background, which is made up from the contributions of many faint unresolved sources, into source flux. This is why this feature does not appear in the results for Method B. While the local background fitting added to Method A goes some ways to alleviating this it is clear that this approach is not completely effective. The other two methods (B and C) do not consider any non-zero background. Simultaneously fitting a solid background under the entire map would almost certainly resolve this issue, however it is not computationally feasible to solve for more than a few hundred sources at once, and hence this is not currently possible. An alternative is to iteratively solve for the background, i.e. fit, and remove, all the known sources in the map (considering no background) and then calculate the mean of the residual map. After one pass this will not give an accurate estimate of the true background, as the source fluxe densities will be underestimated and hence some flux will remain from known sources. However if we repeat this process a number of times, until the mean of the residual map converges, this will give an accurate estimate of the background due to unknown confusing sources. An approach similar to this will likely by used in the next iteration of the HerMES XID algorithm.

\begin{table*}

\caption{Summary of completeness and flux density accuracy for XID methods. In measuring the completeness we consider only sources which are detected at 5$\sigma_{\rm catalogue}$ and pass the quality control thresholds discussed in the text. For the flux density accuracy all sources which are returned by each method are considered. All values in mJy. Catastrophic failures are defined as those that are outside of the 3$\sigma$ range of the best fit Gaussian to the distributions shown in Figure \ref{fig:simerrs}.}
\label{tab:simresults}
\begin{scriptsize}
\begin{tabular}{lllllllllllll}

\hline\hline
\multicolumn{13}{c}{Deep Simulation, 24 $\mu$m$>50 \mu$Jy} \\
\hline
% & \multicolumn{2}{c} {SSX+$p$-stat} & \multicolumn{2}{c} {Linear inversion} & \multicolumn{2}{c} {Linear inversion w/model selection}\\
 & \multicolumn{3}{c}{$\langle$S$_{\rm obs}-$S$_{\rm true}\rangle$}& \multicolumn{3}{c}{RMS } & \multicolumn{3}{c}{Catastrophic failure rate} &\multicolumn{3}{c}{Completeness } \\
 & \multicolumn{3}{c}{(mJy)} & \multicolumn{3}{c}{(3$\sigma$ clipped; mJy)} & \multicolumn{3}{c}{(per cent)} &\multicolumn{3}{c}{($S_{\lambda}[50$ per cent$]$; mJy)} \\
\hline
& $250\,\mu$m & $350\,\mu$m & $500\,\mu$m& $250\,\mu$m & $350\,\mu$m & $500\,\mu$m&$250\,\mu$m & $350\,\mu$m & $500\,\mu$m&$250\,\mu$m & $350\,\mu$m & $500\,\mu$m\\
\hline
Method A (HerMES) &-3.1 & -2.9& -1.8&4.9 & 5.4& 6.5 & 2.4& 2.2 & 9.6 &10.8 &9.8 &26.3\\
Method A' (HerMES,$\rho<0.8$,$\chi^2<5$)& -3.3& -3.1& -2.&4.65 & 4.9& 6.8 & 2.9& 2.1 & 12.9 &11.2 &11.3 &33.3\\
Method B (SSX+$p$-stat) & -0.7& 1.4& 1.7& 7.5 & 9.8 & 9.1 & 9.6 & 2.1 & 1.5 & 39.6 & 47.7& 44.7\\
Method C (Linear Inversion) & -4.3& -3.4& -3.4 &6.3 & 6.8 & 7.8 & 9.5 & 3.7& 2.0 & 20.3&16.8 &16.4\\

\hline
\multicolumn{13}{c}{Shallow Simulation, 24 $\mu$m$>200 \mu$Jy} \\
\hline
Method A (HerMES) &-3.8 &-4.2 & -2.4& 10.1 & 10.8& 14.5 & 2.2 & 2.5 & 2.2 & 24.5 &  26.4 & 38.4\\
Method A' (HerMES,$\rho<0.8$,$\chi^2<5$) &-3.8 & -4.3&-2.5 & 9.25 & 10.3& 14.4 & 2.1. & 2.5 & 2.1 & 24.5 &  26.5 & 38.6\\
Method B (SSX+$p$-stat) &0.4 & 1.5& 3.2& 9.1 & 10.7 & 12.3 & 3.6 & 3.8 & 3.7 & 42. & 46.8 & 53.\\
Method C (Linear Inversion) & -2.9&-4.2 &-4.9 & 9. &10.7 &15.8 &2.1 &4.4 &2.1 &28.7 &27.4 & 34.5  \\

\hline

\end{tabular}
\end{scriptsize}
\end{table*}

\begin{figure}
\includegraphics[scale=0.4]{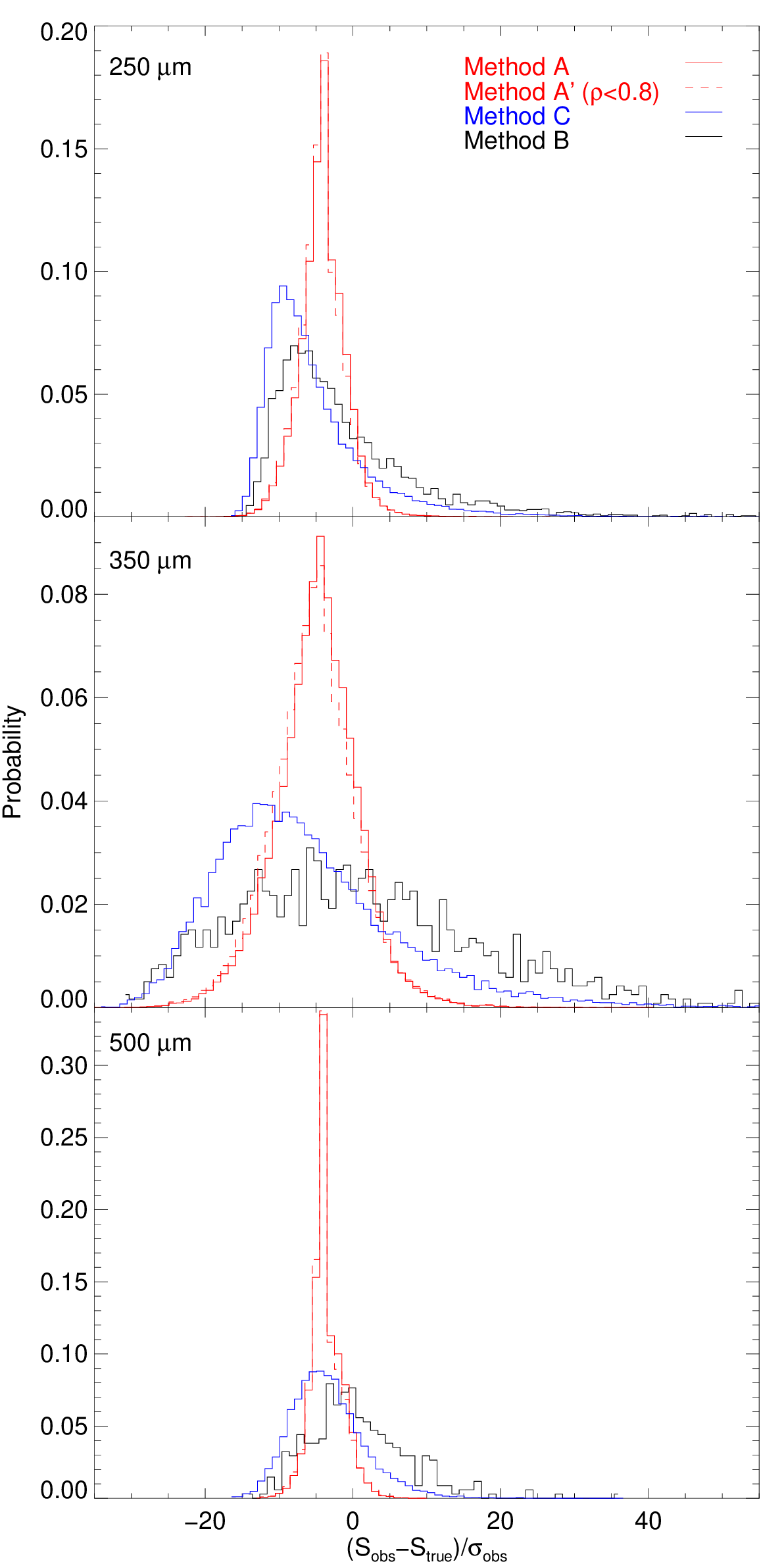}
\caption{Distribution of observed flux density error, normalised by $\sigma_{\rm catalogue}$, for the deep simulation.}
\label{fig:simerrs}
\end{figure}

While high flux density accuracy and completeness are a key aim for any XID method, it is also vital that our proposed method return reliable estimates of the flux density error, as for real applications we will not have knowledge of the true flux density of our sources. In Figure \ref{fig:simerrs} we show the distribution of observed flux density error (i.e. S$_{\rm obs}$-S$_{\rm true}$), normalised by the error estimated by the photometric pipeline for the deep simulation. The first obvious feature of these distributions, as previously discussed in reference to Figures \ref{fig:simcomp} and \ref{fig:simcompshallow}, is that the peak in flux density error distribution is always negative. One side-effect of this systematic negative offset is that it makes the definition of the catastrophic failure rate problematic, if we simply take the number of sources which have abs[$S_{\rm obs}$-$S_{\rm true}$], greater than 3$\sigma_{\rm catalogue}$ then a very large fraction of sources will be considered failures. Thus we take an alternative approach, as we ultimately want to treat our flux densities errors as Gaussian, it makes sense to fit the distributions shown in Figure \ref{fig:simerrs} with a Gaussian, considering the amplitude, mean and sigma as free parameters. Table \ref{tab:distfits} describes the parameters of the best fit Gaussian to each of the distributions shown in Figure \ref{fig:simerrs}.

\begin{table}

\caption{Best fit Gaussian parameters to the normalised flux density error distributions shown in Figure \ref{fig:simerrs} and equivalent distributions for the shallow simulation. All measurements in units of the estimated error $\sigma_{\rm catalogue}$}
\label{tab:distfits}
\begin{tabular}{l|l|l|l|l|l|l}
\hline\hline
 \multicolumn{7}{c}{Deep Simulation}\\
\hline
 & \multicolumn{2}{c}{250$\,\mu$m} & \multicolumn{2}{c}{350$\,\mu$m} & \multicolumn{2}{c}{500$\,\mu$m} \\ 
 & mean & $\sigma$ & mean & $\sigma$ & mean & $\sigma$\\
\hline
Method A & -4.1 & 2.5 & -4.6 & 4.8 & -4.1 & 1.1 \\ 
Method A' & -4.3 & 2.5 & -5.2 & 5.1 & -4.1 & 1\\
Method B & -5.3 & 5.5 & -2.2 & 17 & -0.5 & 6.\\
Method C & -8 & 4.1 & -9.4 & 10.1 & -4.2 & 4.4\\
\hline
 \multicolumn{7}{c}{Shallow Simulation}\\
\hline
Method A & -1.3 & 1.8 & -1.6 & 1.8 & -1.5 & 1.5 \\ 
Method A' & -1.3 & 1.8 & -1.6 & 1.8 & -1.5 & 1.5\\
Method B & -1 & 1.9 & -0.6 & 2.5 & -0.3 & 1.6\\
Method C & -1.7 & 1.8 & -1.8 & 2.1 & -1.5 & 1.7\\
\hline
\end{tabular}
\end{table}

It is clear that for deep observations the quoted catalogue error from the HerMES XID method (Method A) underestimates the true error by a factor of at worst $\sim5$. This is consistent with the values quoted in Table \ref{tab:simresults}, as the typical catalogue error estimated for the deep simulated catalogues is 0.9, 0.7 and 1.5 mJy for the 250, 350 and 500$\,\mu$m bands, respectively. Fortunately the situation is much better once we reach the level of the shallow simulations, where the catalogue errors are consistently within a factor of 2 of the true error. The underlying reason for this discrepency between the true errors and those estimated from the data alone is not completely clear. One possible origin is erroneous fluctuations in the background, which could be eliminated by requiring a smooth background across the entire image, rather than fitting local backgrounds. Another factor will be the incompleteness of the input lists, due to the 24$\,\mu$m flux limit. One puzzling feature is the large variation between the bands. It is worth noting that this variation is quite similar to the variation in input Gaussian noise to the simulations, as quoted in Table \ref{tab:simnoise}. This is suggestive of a hard limit to the flux density error, either from the factors listed above, or simply noise introduced from the deblending of confused faint sources.

One thing which is clear is that the other potential XID methods are significantly worse at accurately estimating the flux density error. While Method A shows a very Gaussian distribution of normalised flux densities errors, the other methods have a long tail to very large values. To quantify this we define the catastrophic failure rate as the fraction of sources which appear at abs[$S_{\rm obs}$-$S_{\rm true}$]$>3\times\sigma_{fit}$, where $\sigma_{fit}$ is the best fit value derived for a specific SPIRE band and method in Table \ref{tab:distfits}. At 250$\,\mu$m it is clear that Method A returns a highly Gaussian error distribution, with only $\sim 2$ per cent falling outside the 3$\sigma_{fit}$ range. The other methods have a much higher catastrophic failure rate at 250$\,\mu$m, approaching $\sim10$ per cent. At the other SPIRE bands, the 350$\,\mu$m distributions are well described by a Gaussian for all methods, but at 500$\,\mu$m it appears that there are a significant fraction of catastrophic failures produced by Method A. While these failures are still quite reliable compared to the very large errors returned by the other methods, it is worth commenting on this non-Gaussian element to the distribution. This is likely an artifact of the model selection being performed at 250$\,\mu$m only, as high redshift 500$\,\mu$m `peaking' sources will appear faintly in the 250$\,\mu$m maps and hence are likely to be missing from the input list at 250$\,\mu$m. In these cases the 500$\,\mu$m flux will be erroneously assigned to the neighbouring 250$\,\mu$m bright source. 

Interestingly the flux density errors for Method A and A' are in good agreement with the measured confusion noise limit from Nguyen et al. (2010). Thus it is clear that our method is able to probe flux densities close to, if not below, the confusion noise. This is particularly noteworthy when considering that the systematic negative offset in the flux densities, due to issues with the background fitting, is a large contributer to this noise.

As one of the key science goals of SPIRE surveys will be investigations of far-IR SEDs and their evolution, we need to understand not only the quality of the monochromatic SPIRE flux densities, but also any correlations between the bands. To investigate this in our simulated data set we look for correlated errors in the flux density accuracy of the deep simulation results.

\begin{figure*}
\includegraphics[scale=0.9]{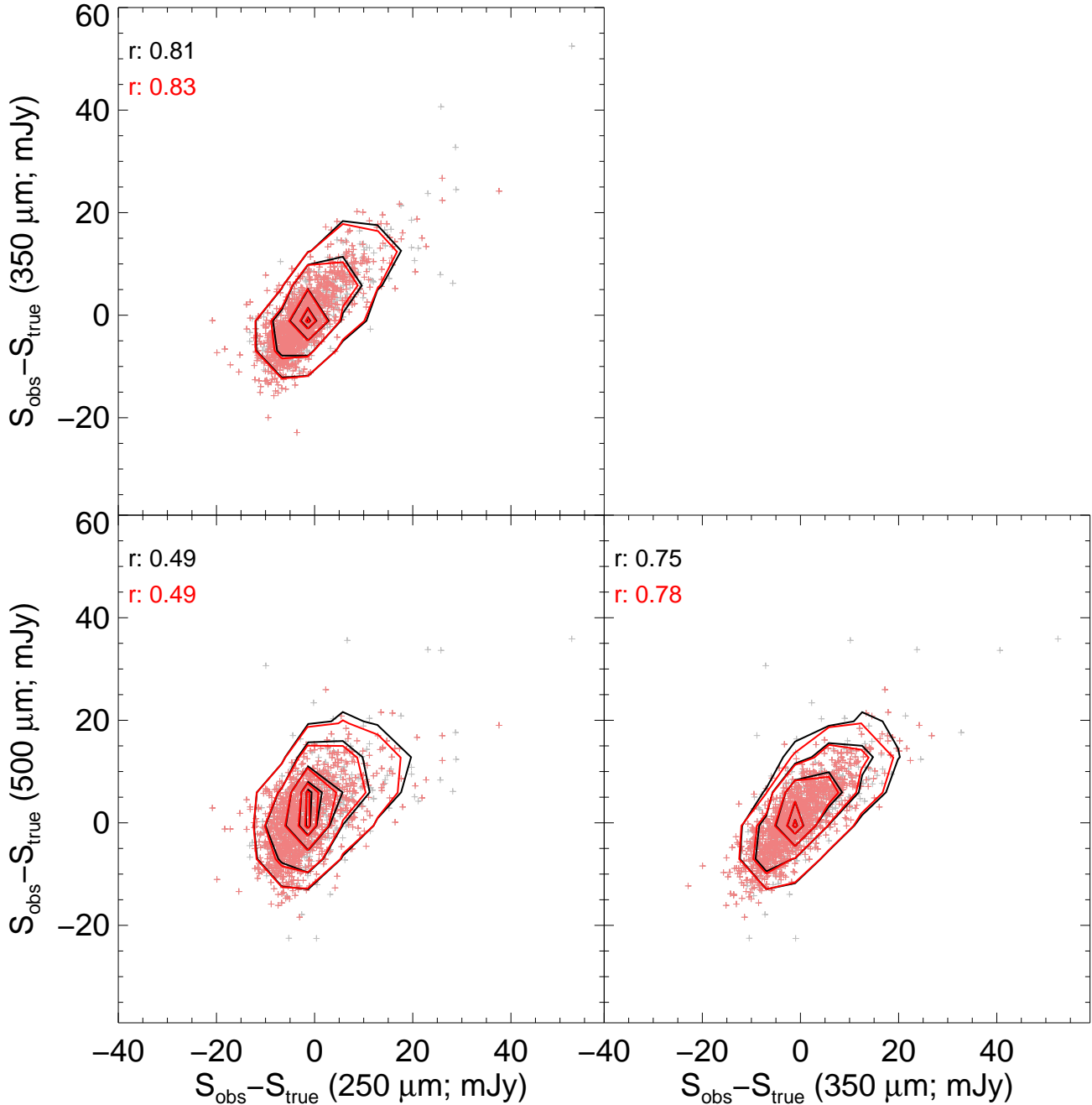}
\caption{Correlations in the flux density errors found in the deep simulations. All 5$\sigma$ sources are shown in grey (black contours), while those sources that also have $\rho<0.8$ are shown in pink (red contours). The Pearson correlation coefficient ($r$) is quoted in the top left corner of each panel.   }
\label{fig:simcor}
\end{figure*}

Figure \ref{fig:simcor} shows the relationship between the flux density error ($S_{\rm obs}-S_{\rm true}$) in the three bands for the deep simulations. It is clear that the flux density errors show a strong linear correlation. Quantifying these correlations with the Pearson correlation coefficient ($r$) shows that the 250 to 350$\,\mu$m and 350 to 500$\,\mu$m flux density errors are strongly correlated ($r\sim0.8$), while the 250 to 500$\,\mu$m flux density errors show somewhat weaker correlations ($r=0.5$). Performing similar tests on the shallow simulation and other XID methods gives similarly strong correlations.

Although the peculiarities of the XID algorithms could be partially responsible for these correlations, the underlying origin must be the effect of unknown, or poorly deblended, close neighbours. While our method is designed to optimally deblend sources in the input list, this can never be perfectly achieved without perfect input lits. Given this it is unlikely that modifications can be made to the XID algorithm to remove these correlations. One thing to note is that the correlations are dependent not only on the areal density of sources, but also on the far-IR colours. Since the FC08 mock skies include only a limited range of SED types and SED/flux density independent clustering it is reasonable to assume that the amplitude of these correlations will be weaker in the real data. %However, it is not possible to quantify the correlated errors between the bands without using a more sophisticated source fitting algorithm which considers all three bands simultaneously.

\begin{figure}
\includegraphics[scale=0.45]{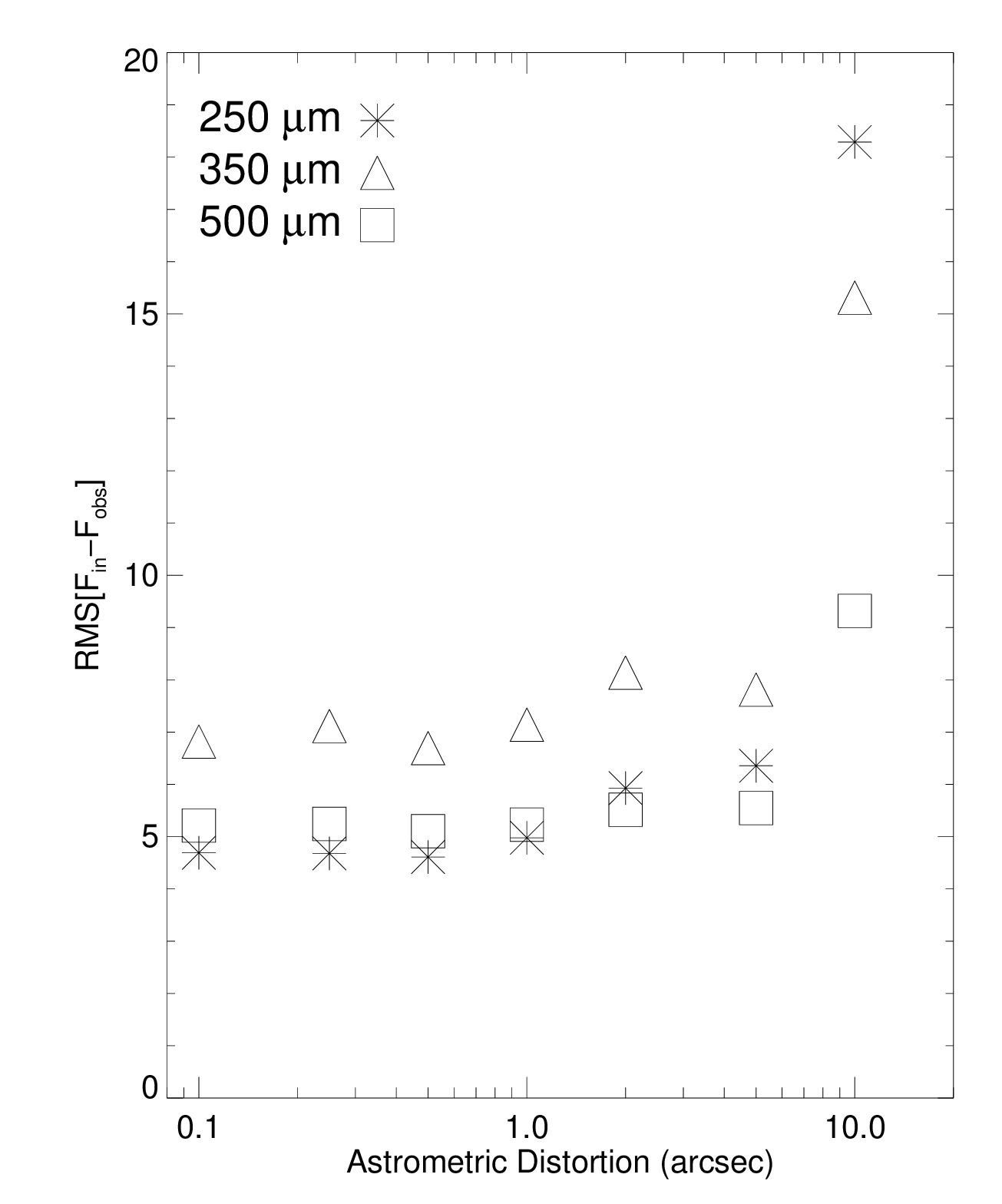}
\caption{Effect of astrometric errors on the flux density accuracy of the HerMES XID algorithm (Method A). Gaussian distributed distortions are added to the input positions of the deep simulation described in Table \ref{tab:simnoise}. It can be seen that the accuracy of the flux density estimates is insenstive to astrometric errors of $<1-2$ arcsec.}
\label{fig:asterrs}
\end{figure}

Finally, while in the simulations described above the input positions here have no astrometric errors in real applications the input lists and SPIRE images will have errors on the order of $\sim 0.1-0.5$ arcsec. Thus it is worth considering the effects of astrometric distortions on our simulated dataset. To achieve this normally distributed random astrometric errors are added to the input positions and the XID process repeated. Here we only consider the HerMES XID algorithm (Method A). Figure \ref{fig:asterrs} shows the result of adding errors on the scale of 0.1--10 arcsec to our input list.  It can be seen that the accuracy of the flux density estimates is insenstive to astrometric errors of $<1-2$ arcsec.

\section{Testing on Real Data}\label{sec:xidvsscat}

While it is useful to assess the completeness and flux density accuracy of our method on totally artificial maps, we can also calculate these metrics for the real data by injection of mock sources into our observed maps. This has the advantage of reproducing the true noise properties of the data, as well as highlighting the confusion noise in the presence of angular clustering.

\begin{figure*}
\includegraphics[scale=0.8]{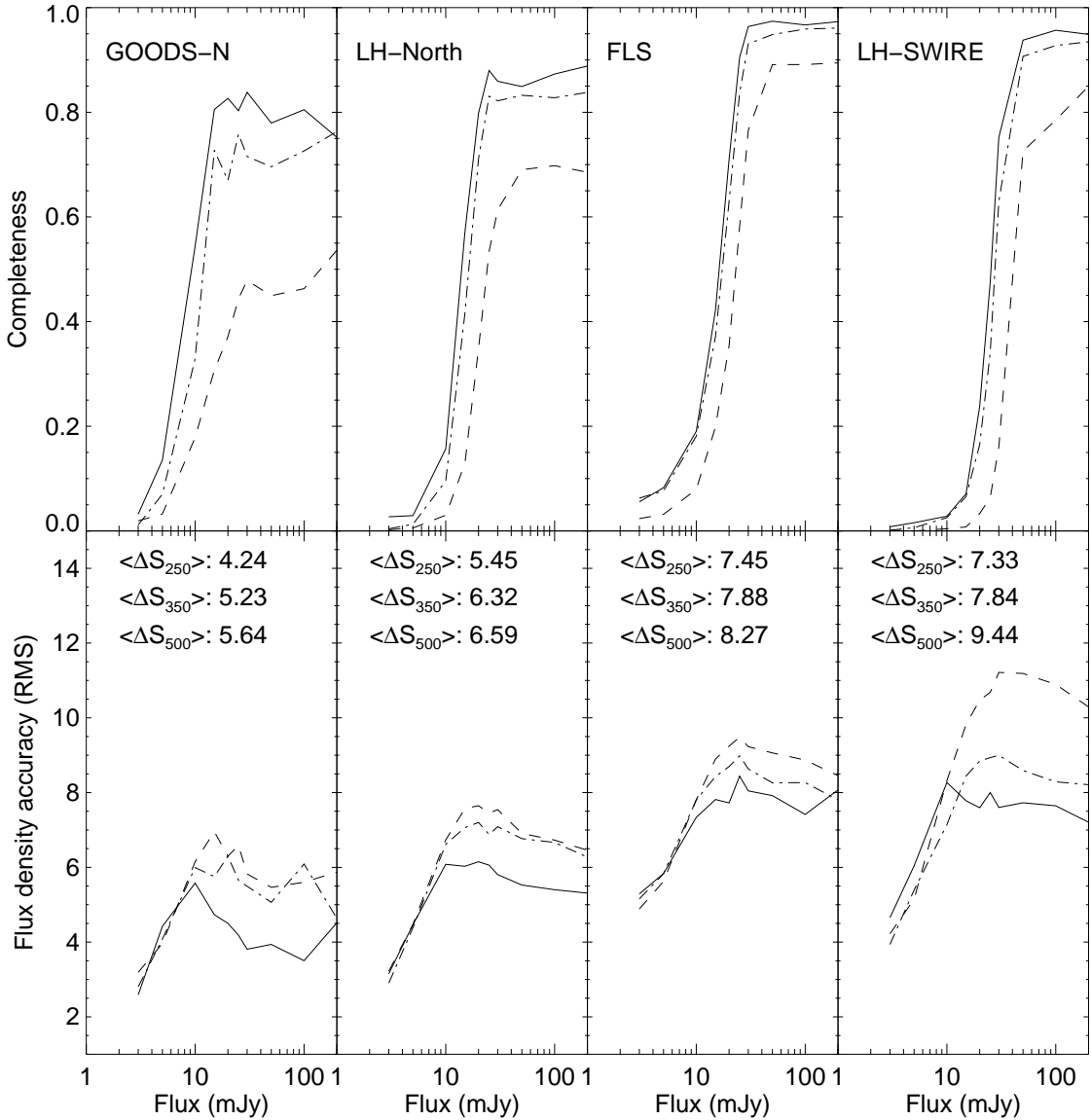}
\caption{Completeness (top) and flux density accuracy (bottom) determined by injection of mock sources into our observed maps. The completeness is defined as the ratio of the number of sources recovered at $>5\sigma$ and $\rho<0.8$ to the number of input positions.  Flux density accuracy is defined as the RMS of the input--output flux density. In calculating the recovered flux density accuracy all input positions recovered with $\rho<0.8$ are considered. Sources are injected one at a time so as to avoid increasing the source confusion. In each panel the results for the $250\,\mu$m (solid line), $350\,\mu$m (dot-dashed line) and $500\,\mu$m (dashed line) bands are shown. Mean flux density error for each band is shown in the top right corner of each of the lower panels.}
\label{fig:injectresults}
\end{figure*}

As our maps are already heavily affected by confusion, we only inject one source at a time into the map, and then run the XID source extraction algorithm, taking the input position of the mock source and the neighbouring $24\,\mu$m sources into account. For each SDP field we inject mock sources with flux densities in the range 3--200~mJy at random positions. Test positions outside of the $24\,\mu$m coverage are not considered. To maintain consistency with the properties of the real $24\,\mu$m input catalogues, test positions within 3 arcsec of an existing $24\,\mu$m source are also excluded, as was done with the fully articifial simulations. As a result the total number of test positions is $\sim 3000$--5000 per field, with 300--500 per test flux density. Figure \ref{fig:injectresults} shows the completeness and flux accuracy determined by this method.

It can be seen that the completeness never reaches 100 per cent in any field. The values rise sharply from faint flux densities and then plateau at a quasi-constant value above a certain flux density level. This is due to the effect of the $\rho<0.8$ criteria. Somewhat counter-intuitively, this is a bigger problem in the fields with deeper SPIRE/MIPS $24\,\mu$m data. The reason for this is simple; the input source density is much higher in the deep fields and, as we assume no prior on the SPIRE flux density, this affects all flux densities equally. If the $\rho<0.8$ criteria is removed the residual $\sim 20$--50 per cent incompleteness in the deep fields is recovered, but at the expense of flux density accuracy. For sources with $\rho<0.8$ in the GOODS-N field the 1$\sigma$ flux density error is 4.24, 5.23 and 5.64 mJy for the 250, 350 and $500\,\mu$m bands, respectively. For sources with $\rho>0.8$ the comparable values are 6.3, 5.9, and 6.9 mJy, an increase of $\sim10$--50 per cent. %More troubling, the catastrophic failure rate, i.e., the fraction of sources with $S_{in}-$$S_{obs}>3\sigma$, is 1.3 per cent, 1.1 per cent, and 0.8 per cent for the $\rho<0.8$ sources and 1.4 per cent, 2.3 per cent and 3.6 per cent for the $\rho>0.8$ sources.% [SHOULD I PUT ALL THESE NUMBERS IN A TABLE?]. 

Encouragingly, the completeness and flux density accuracy derived from source injection agrees reasonably well with the numbers for comparable simulations (Table \ref{tab:simresults}). Two small exceptions to this are the completeness at 500 $\mu$m in the deep simulation/GOODS-N and the error in the 500 $\mu$m flux density in the shallow simulation/Lockman-SWIRE. In the first instance the observed completeness in the real maps is slightly lower than that found in the simulations. The origin of this is not clear, but it is likely caused by slight differences in the input list. The simulations use a hard S$_{24}>50~\mu$Jy cut, while the real GOODS-N catalogue is cut at S$_{24}>5\sigma$, which includes many sources fainter than S$_{24}=50~\mu$Jy and hence has a higher surface density, leading to more degenerate solutions. Another possible explanation is that real 500 $\mu$m sources are more strongly clustered than those in the simulation. The second issue, the difference in the flux density error at 500 $\mu$m between the shallow simulation and Lockman-SWIRE, likely originates from differences in how the 24 and 250 $\mu$m selection affects the real and simulated datasets. Specifically, larger errors would be expected if the simulations predict a higher level of incompleteness at 500 $\mu$m due to the 24 and/or 250 $\mu$m selections. These problems aside the otherwise good agreement reinforces the notion that our simulations are a realistic recreation of the {\it Herschel} data. 

These results highlight the effectiveness of our method to recover faint sources in highly confused maps, Nguyen et al. (2010) estimate the confusion noise in SPIRE imaging to be 5.8, 6.3 and 6.8 mJy at 250, 350 and 500$\,\mu$m, respectively. It is clear that in our deepest fields, where instrumental noise is insignificant we are able to go significantly below this limit. Taking the 1$\sigma$ flux density error quoted above for sources in GOODS-N with $\rho<0.8$ it is clear our methods are able to reduce the effect of confusion noise by a factor of $\sim$20--30 per cent.

\begin{table*}
\caption{Completeness estimates (50 per cent) for XID and SCAT+$p$-stat catalogues for real observations of SDP fields. Completeness is estimated via both injection of sources into the map, and by comparing the number density of sources in the resulting catalogues to the best estimate of the true source density from Oliver et al. (2010b).}
\label{tab:obscomp}
\begin{footnotesize}
\begin{tabular}{l|ccc|ccc|ccc}
\hline\hline
 & \multicolumn{3}{c}{$S_{250}[50$ per cent$]$} & \multicolumn{3}{c}{$S_{350}$[50 per cent]} & \multicolumn{3}{c}{$S_{500}$[50 per cent]}\\
& \multicolumn{3}{c}{(mJy)}&\multicolumn{3}{c}{(mJy)}&\multicolumn{3}{c}{(mJy)}\\
\hline
 & \multicolumn{1}{c}{Source Injection} & \multicolumn{2}{c}{Counts} & \multicolumn{1}{c}{Source Injection} & \multicolumn{2}{c}{Counts} & \multicolumn{1}{c}{Source Injection} & \multicolumn{2}{c}{Counts} \\
Field &XID&XID&SCAT+$p$-stat &XID &XID & SCAT+$p$-stat&XID&XID& SCAT+$p$-stat\\

\hline
GOODS-N &9.5  & -- & -- & 12.1 &-- & -- & n/a  & -- & -- \\
LH-North &14.1  & 13.3 & 26.7 & 16.4 &23.8 & 30.5 &24.1  &23.6 & 31.5\\
FLS & 16.4 & 21.6 & 23.3 & 17.5 &21.6 & 23.3 &23.2  &22.9 & 25.5\\
LH-SWIRE & 25.4 & 27. & 36.4 & 27.8 & 26.7 & 35.6 & 42. &36.4 & 44.4 \\
\hline
\end{tabular}
\end{footnotesize}
\end{table*}

To investigate possible systematics in our photometry we compare the XID catalogues to those generated using a combination of source detection and extraction, via Sussextractor and $p$-statistic methods (i.e. method B from Section \ref{sec:simresults}), to match the resulting source lists with existing $24\,\mu$m catalogues.

Sussextractor source lists are provided for each SPIRE band by SCAT (Smith et al. 2010). These source lists contain all SPIRE sources detected in the maps at a significance of greater than $3\sigma$. The monochromatic SPIRE source lists are then matched to the same $24\,\mu$m catalogues used as an input to XID algorithm. The matching is performed by finding potential counterparts within a search radius of 10 arcsec, 14 arcsec, and 20 arcsec for the 250, 350 and $500\,\mu$m bands respectively. For each of these potential IDs we calculate the $p$-statistic. The uncertainty of the SPIRE position is calculated using Equation B8 of Ivison et al. (2007). All IDs with $p<0.1$ are considered. A complete sample is constructed by taking the best ID with $p<0.1$ for each SPIRE source. Alternatively a `clean' sample is constructed by taking only those cases where the separation is less than $0.6\times$FWHM, and there is only one potential ID with $p<0.1$.

\begin{figure}

\includegraphics[scale=0.6]{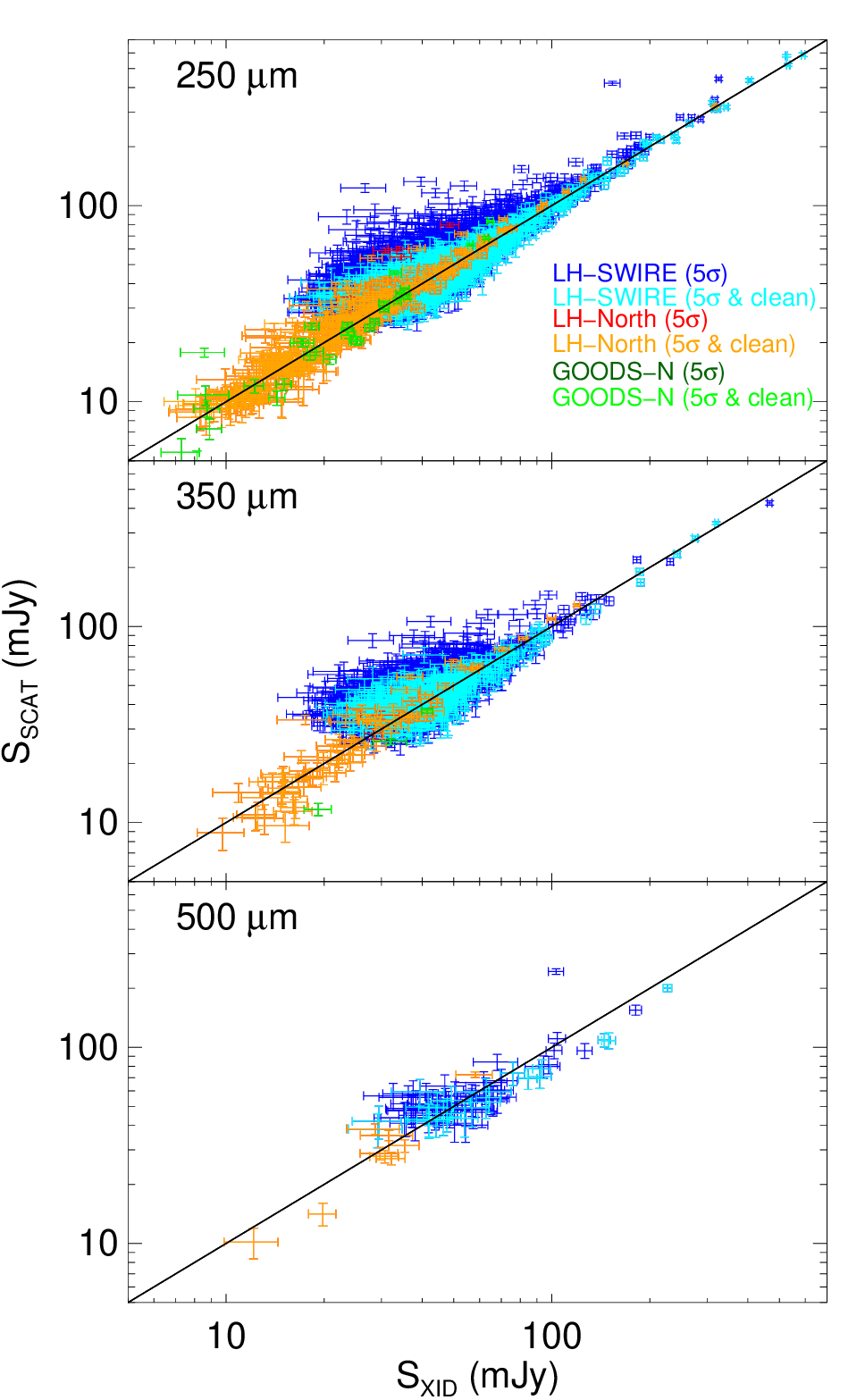}
\caption{Comparison of flux densities from XID catalogue to those from the SCAT SussExtractor-derived source catalogues in the Lockman Hole SWIRE, Lockman Hole North and GOODS-N fields. SCAT sources are matched to the $24\,\mu$m sources via the $p$-statistic. Sources present in both catalogues at 5$\sigma$ are presented, as well as a ``clean'' sample where $p<0.1$, seperation$<0.6\times$FWHM$_{\rm SPIRE}$, and there are no alternative IDs with $p<0.1$. XID fluxes are also required to have $\rho<0.8$ and $\chi^2<5$. Good agreement can be seen between the XID and SCAT flux densities for ``clean'' sources. This suggests that any discrepencies between SCAT and XID are solely due to issues with source blending. }
\label{fig:xidflux densities}
\end{figure}

Figure \ref{fig:xidflux densities} compares the flux density estimates for sources in the LH-SWIRE, LH-North fields and GOODS-N from the XID catalogues and the SCAT+$p$-stat listings. Only those sources which are in common and are found at greater than $5\sigma$ in both catalogues are presented. The FLS field is omitted for clarity. While there is a large scatter between the two estimates for all sources, a good agreement can be seen for the `clean' ones. The bulk of the sources which are discrepant between the two catalogues can be found above the one-to-one line in Figure \ref{fig:xidflux densities}, i.e. $S_{\rm XID}<S_{\rm SCAT}$. This is a natural consequence of the XID algorithm considering all known sources simultaneously, and thus deblending confused cases into their individual 24$\,\mu$m detected components.

\begin{figure}

\includegraphics[scale=0.45]{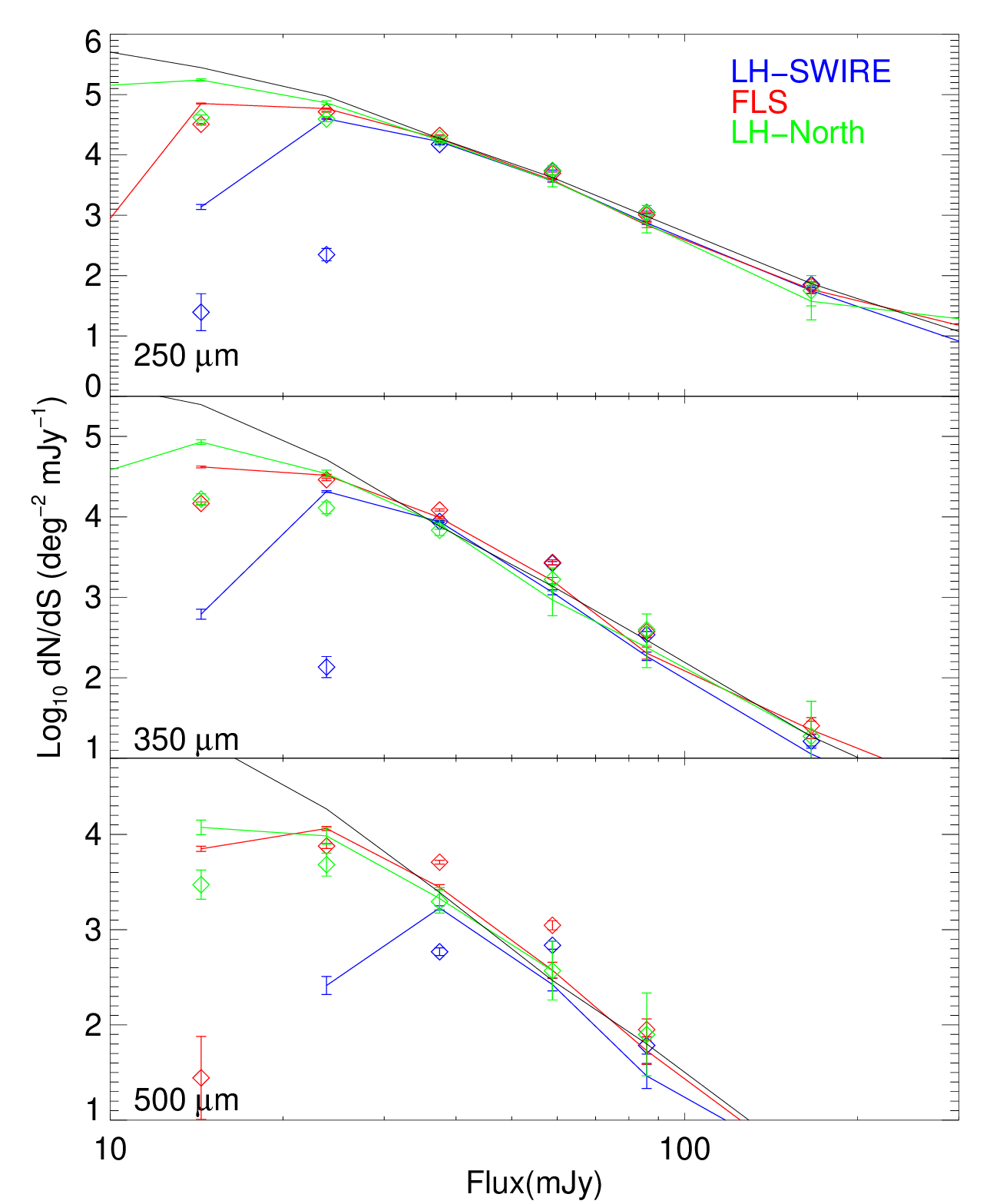}
\caption{Differential number density of sources in the XID catalogues and SCAT v3 release catalogues. The black line is the current best estimate of true source density from Oliver et al. (2010b). The solid lines are XID, open symbols are SCAT in all panels.}
\label{fig:xidcnts}
\end{figure}

As a final cross-check of the completeness estimates we compare the raw differential number density of sources found in both the XID and SCAT+$p$-stat catalogues to the best estimates of the source densities from Oliver et al. (2010b). Figure \ref{fig:xidcnts} shows the differential number density of sources in our XID and SCAT+$p$-stat catalogues, in the LH-SWIRE, FLS, and LH-North fields. GOODS-N observations are excluded as the number of sources detected is too small for this comparison to be useful. Encouragingly, at bright flux densities (i.e. $>50$ mJy), both the XID and SCAT+$p$-stat catalogues show reasonable agreement with Oliver et al. (2010b), although cosmic variance introduces a large scatter at the highest flux densities. Both the XID and SCAT+$p$-stat are seen to be incomplete at faint flux densities, although in each band the XID catalogue is significantly more complete at flux densities $\sim$ 20--30 mJy. Taking the Oliver et al. result to represent the total number of sources, Table \ref{tab:obscomp} quotes the XID and SCAT+$p$-stat catalogue 50 per cent completeness levels. These values are in good overall agreement with the completeness estimates found via simulations and source injection.

\section{The effect of incomplete 24 $\mu$\lowercase{m} input lists}\label{sec:mipsincompleteness}
We have shown that the use of existing $24\,\mu$m source lists as a prior input to the source extraction process is beneficial in terms of flux density accuracy and completeness. However this methodology introduces a clear bias, in that we are restricted to only those sources which are sufficiently bright at both $24\,\mu$m {\it and} SPIRE wavelengths. 

One way to estimate this incompleteness is to again use the mock catalogues. Again turning to the FC08 mocks we can estimate the fraction of sources which would be present in our SPIRE images, but below the limit of the overlapping 24$\,\mu$m imaging. For the SPIRE bands we use the 50 per cent completeness limits quoted in Table \ref{tab:obscomp}, while for the 24$\,\mu$m flux density limit we use the values quoted in Table \ref{tab:sdpObs}. As the GOODS-N field never reaches 50 per cent completeness we use the value from the deep simulation presented in Table \ref{tab:simresults}. The fraction of sources missing due to the 24$\,\mu$m limit in the FC08 mock catalogues is given in Table \ref{tab:mipsIncompleteness}

However relying on mock catalogues to describe this incompleteness is unsatisfactory, as it is very sensitive to the underlying SED distribution of sources, a known weakness of mock catalogues based on emphirical fits to the observed monochromatic number density of sources.

In order to properly determine what additional incompleteness this introduces would require precise measurements of the bivariate number density, i.e. the areal number density of sources as a function of both $24\,\mu$m and SPIRE flux density. While that analysis is beyond the scope of this work, we can roughly estimate the lower limit to this incompleteness in our fields by making use of the multi-tiered nature of HerMES. Specifically we can use our observations in GOODS-N, which contains both the deepest SPIRE imaging and deepest 24$\,\mu$m catalogues available, to determine the number of sources which would appear in the fields with shallower SPIRE data, if similar quality $24\,\mu$m input catalogues were available.

\begin{figure}
\includegraphics[scale=0.6]{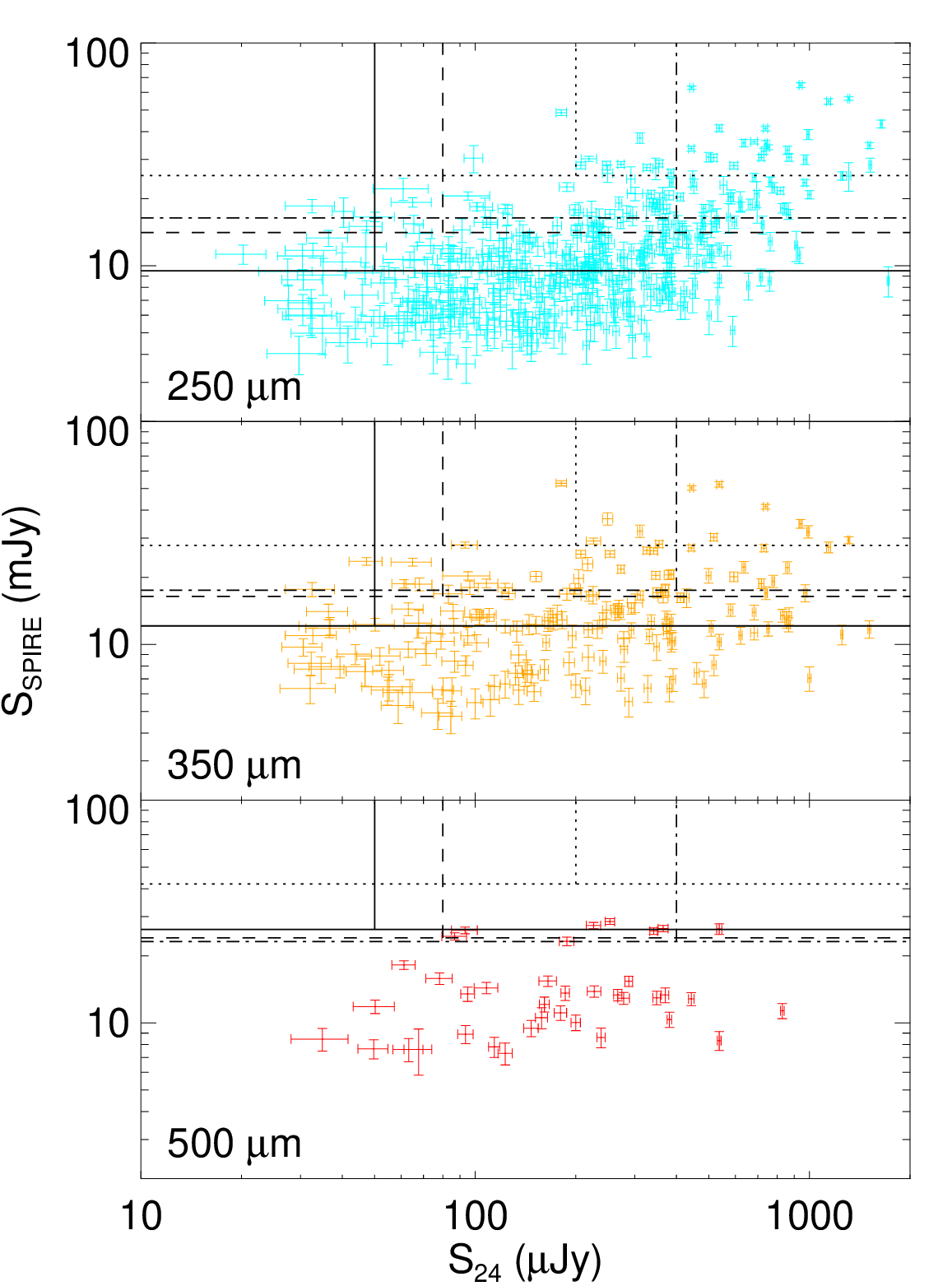}
\caption{SPIRE vs. 24$\,\mu$m flux density for sources in GOODS-N. The vertical lines indicate the depth of 24$\,\mu$m imaging in each SDP field, while the horizontal lines indicate the 50 per cent completeness level of our SPIRE catalogues from the analysis presented in Section \ref{sec:xidvsscat}. Fields are; GOODS-N (solid), Lockman-North (dashed), FLS (dot-dashed), Lockman-SWIRE (dotted).}
\label{fig:goodsSP24col}

\end{figure}

\begin{figure}

\includegraphics[scale=0.45]{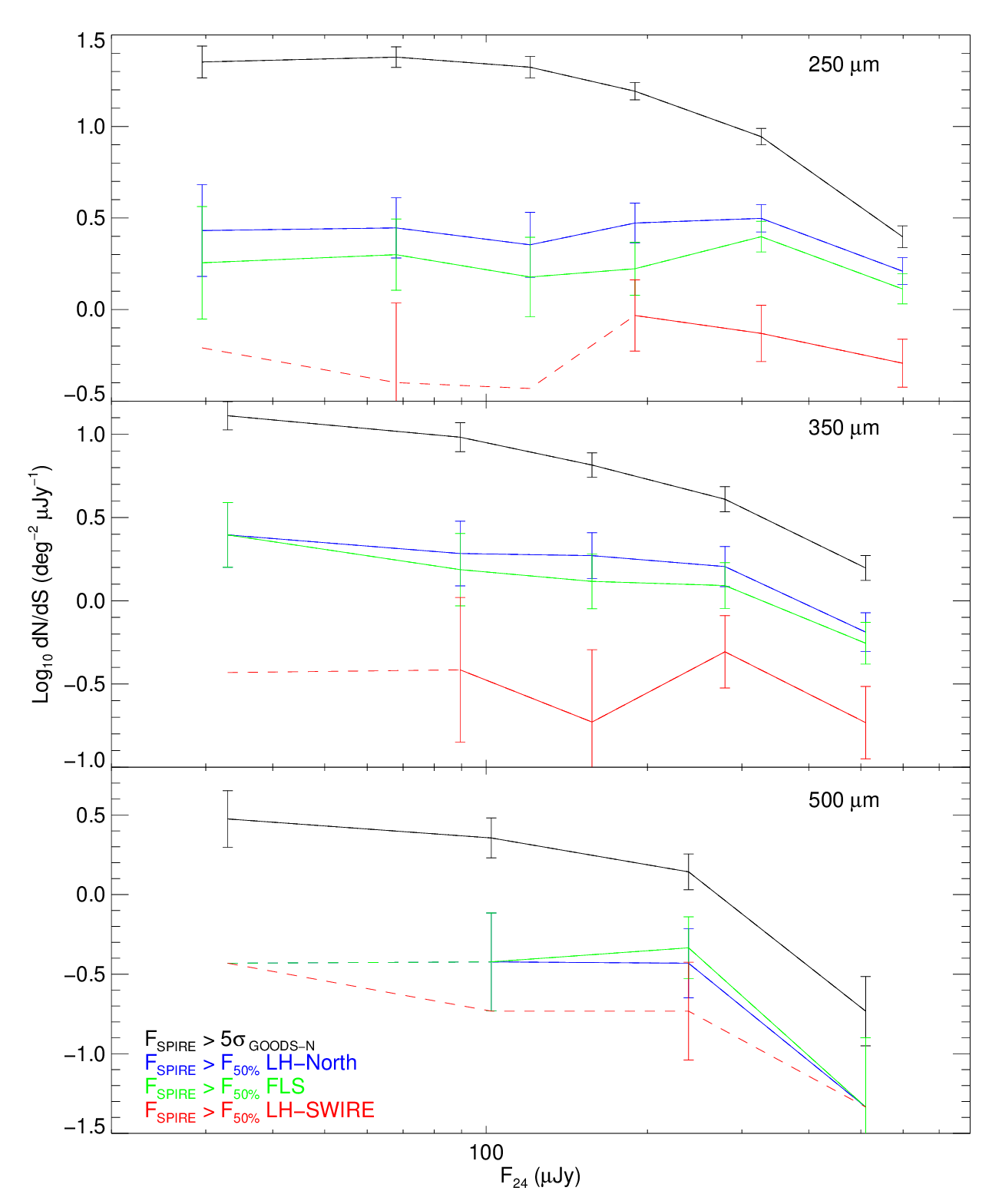}
\caption{Differential number density of SPIRE sources in the GOODS-N field as a function of $24\,\mu$m flux density. The curves have been corrected for incompleteness in the $24\,\mu$m catalogue, but not the SPIRE incompleteness. The three panels show the results for the $250\,\mu$m (top), $350\,\mu$m (middle) and $500\,\mu$m (bottom) band. The black line in each panel shows our best estimate of the differential number density of sources detected in the GOODS-N field as a function of $24\,\mu$m flux density. The other lines show the effect of imposing the SPIRE 50 per cent completeness limit on the GOODS-N catalogue. The dashed line indicates the number density which would be quoted if one source was observed in that bin.}
\label{fig:mipsDN}
\end{figure}

Figure \ref{fig:goodsSP24col} shows the 24$\,\mu$m vs SPIRE band flux density for 5$\sigma$ sources observed in GOODS-N, while Figure \ref{fig:mipsDN} shows our best estimate of the differential number density of SPIRE sources as a function of $24\,\mu$m flux density derived from this data. The densities have been corrected for incompleteness in the $24\,\mu$m input catalogue using the results presented in Magnelli et al. (2009).  We do not correct for SPIRE incompleteness as we wish to estimate how many sources are missing from our catalogues due to solely the $24\,\mu$m flux limits. GOODS-N sources which are 5$\sigma$ in the relevant SPIRE band are considered, with no cut on $\rho$. To replicate the conditions found in our other fields we impose artificial SPIRE flux limits on the GOODS-N data. For each combination of SPIRE band and field we impose the 50 per cent completeness limit found via source injection quoted in Table \ref{tab:obscomp}. To find a robust estimate to the total number of sources missing from our shallower SPIRE observations we integrate Figure \ref{fig:mipsDN} from zero to the quoted $24\,\mu$m limit given in Table \ref{tab:sdpObs}. Below $S_{24}=20 \mu$Jy, or in cases where no sources are observed, we assume the differential density remains constant from zero to the last measured value. Table \ref{tab:mipsIncompleteness} summarises the results of these calculations. Given the very small area covered by GOODS-N, It should be noted that all of these values are subject to large uncertainties, especially at 500$\,\mu$m where the number of bright sources found in GOODS-N is very small.  

\begin{table*}
\caption{Upper limit to the incompleteness in our SDP fields due to $24\,\mu$m flux limits, based on analysis of FC08 mock catalogues, and SDP observations of GOODS-N.}
\label{tab:mipsIncompleteness}
\begin{tabular}{l|l|l|l|l|l|l|l|l|l}
\hline
Field & \multicolumn{3}{c} {$250\,\mu$m} & \multicolumn{3}{c} {$350\,\mu$m} & \multicolumn{3}{c} {$500\,\mu$m} \\
\hline
 & \multicolumn{2}{c}{real} & mock & \multicolumn{2}{c}{real} & mock & \multicolumn{2}{c}{real} & mock \\
 & N$_{\rm missing}$&  per cent & per cent  & N$_{\rm missing}$ &  per cent  &per cent  &N$_{\rm missing}$ &  per cent &per cent   \\
\hline\hline
GOODS-N & -- & -- & 1 & -- & -- & 3 & -- & -- & $<1$\\
LH-North & $80\pm30$ & $10\pm4$ & 1 &$60\pm30$ & $17\pm7$ & 3 & $10\pm4$ & $28\pm12$ & 1\\
FLS & $2900\pm1200$ & $40\pm16$ & 50 & $2300\pm900$ & $50\pm20$ & 30 & $600\pm200$ & $60\pm20$ & 17\\
LH-SWIRE & $2000\pm1000$ & $20\pm10$ & 2 & $800\pm300$ & $20\pm10$ & 5 & $300\pm300$ & $70\pm70$ & 7\\
\end{tabular}
\end{table*}

Encouragingly in both the LH-SWIRE and LH-North field at 250 and $350\,\mu$m we appear to only be missing an additional $\sim20$ per cent of sources due to the $24\,\mu$m depth. The shallow nature of the $24\,\mu$m imaging in FLS means we are missing a significant number of sources in this field, although the bulk of these will be at relatively faint fluxes ($<30$ mJy). At $500\,\mu$m all of the fields potentially suffer from a high degree of additional incompleteness due to the $24\,\mu$m limits. This is understandable, as the strong negative $k$-correction with increasing redshift at $500\,\mu$m should result in a population of high-$z$ $500\,\mu$m bright, $24\,\mu$m faint sources which would not be found via the methodology presented here. 

\section{Future Work}\label{sec:future}

As discussed in Section \ref{sec:HermesXID} the algorithm and catalogues described here represent the first attempt to produce robust XIDs for SPIRE sources, and hence many avenues for improvement are possible in terms of both flux density accuracy and completeness. Some clear improvements have already been discussed above. Specifically in these area:
\begin{itemize}
\item  Perform the model selection stage on all three SPIRE bands, and possibly other MIPS and PACS data, simultaneously.
\item  Introduce flux density priors based on SED fitting.
\item  Improve the process of background estimation and removal
\item  Use an iterative process to recover faint sources missing from our $24\,\mu$m input list.
\item  Obtain accurate estimates of the true errors on our flux densities.
\end{itemize}

Of these the final one, accurate estimation of the errors, is arguably the most critical. It is clear from the accuracy metrics presented in Sections \ref{sec:simresults} and \ref{sec:xidvsscat} that our flux density errors are underestimates of the true variance in our measurements. If the true variance, and covariance, of each flux density estimate could be obtained the use of crude `flags' for selecting robust sources, such as the $\rho$, purity, and local $24\,\mu$m source density, would no longer be necessary. 

One way to more accurately estimate the flux density errors would be to make use of Monte-Carlo Markov Chain (MCMC) methods to perform the linear inversion. This would have many advantages: a MCMC approach would map out the true posterior probability density for not only the source flux density variances, but also the covariance. Additionally, a MCMC approach offers the natural inclusion of `non-linear' prior knowledge on the solutions, such as smooth SED and background constraints. Preliminary testing of a hybrid MCMC method, which makes use of Hamiltonian dynamics to draw samples, on the simulated data presented in Section \ref{sec:simresults} has shown that for typical segment sizes containing $<100$ sources, MCMC chains of length $\sim 10^6$ can robustly recover the true variance in the flux density estimates, although with some loss of precision in the flux density estimate. Further testing with this approach is needed to determine if MCMC based methods can return the best results in terms of both precision and robust error estimation.

\section{Conclusions}
We have presented a new technique for producing associations between astronomical observations at different wavelengths. This method is optimised for use on {\it Herschel} SPIRE imaging in the presence of deep $24\,\mu$m catalogues from {\it Spitzer}. This technique has been used to produce XID catalogues for the HerMES SDP fields. Thorough testing is performed on simulated and real data sets for both our new method, and two existing XID methods. Compared to a more traditional approach of source detection and catalogue-based cross identification, our map-based approach is found to give significantly greater accuracy in the flux density and recovers a much larger fraction of faint SPIRE sources. When compared to the Sussextractor derived source catalogues of Smith et al. (2010) we find good agreement between flux density estimates, for those sources considered to be `unconfused'. We find that the use of the $24\,\mu$m prior input list can introduce an additional incompletness which is strongly dependant on the relative depth of the existing 24$\,\mu$m data to our SPIRE data. From the combination of deep SPIRE and {\it Spitzer} 24$\,\mu$m observations in GOODS-N  we estimate an incompleteness due to the 24$\,\mu$m limit in the other SDP fields of $\sim 20$ per cent at 250 $\mu$m, increasing to $\sim 40$ per cent at 500 $\mu$m. However this incompleteness is dominated by the faintest SPIRE sources (i.e. less than $30$--$40$~mJy), and we can be confident our catalogues are complete at bright fluxes.

\section*{Acknowledgements}
We thank the anonymous referee for many suggestions which greatly enhanced the clarity of this paper.\\

The SPIRE Consortium includes participants from eight countries (Canada, China, France, Italy, Spain, Sweden, UK, USA).  The following institutes have provided hardware and software elements to the instrument programme: Cardiff University, UK;  Commissariat  à  l'Énergie Atomique (CEA), Saclay, France; CEA, Grenoble, France; Imperial College, London, UK; Instituto de Astrofisica de Canarias (IAC), Tenerife, Spain; Infrared Processing and Analysis Centre (IPAC), Pasadena, USA; Istituto di Fisica dello Spazio Interplanetario (IFSI), Rome, Italy; University College London's Mullard Space Science Laboratory (MSSL), Surrey, UK; NASA Goddard Space Flight Centre (GSFC), Maryland, USA; NASA Jet Propulsion Laboratory (JPL) and Caltech, Pasadena, USA; National Astronomical Observatories, Chinese Academy of Sciences (NAOC), Beijing, China; Observatoire Astronomique de Marseille Provence (OAMP), France; Rutherford Appleton Laboratory (RAL), Oxfordshire, UK; Stockholm Observatory, Sweden; UK Astronomy Technology Centre (UK ATC) Edinburgh; University of Colorado, USA; University of Lethbridge, Canada; University of Padua, Italy; and the University of Sussex, UK.  Funding for SPIRE has been provided by the national agencies of the participating countries and by internal institute funding: the Canadian Space Agency (CSA); NAOC in China; Centre National d'Études Spatiales (CNES), Centre National de la Recherche Scientifique (CNRS), and CEA in France; Agenzia Spaziale Italiana (ASI) in Italy; Ministerio de Educación y Ciencia (MEC) in Spain, Stockholm Observatory in Sweden; the Science and Technology Facilities Council (STFC) in the UK; and NASA in the USA.  Additional funding support for some instrument activities has been provided by ESA.

\label{lastpage}


\begin{thebibliography}{99}
%\bibliography{lit}
%\bibliographystyle{mn2e}
\bibitem[Akaike(1974)]{1974ITAC...19..716A} Akaike, H.\ 1974, IEEE 
Transactions on Automatic Control, 19, 716 
\bibitem[Austermann et al.(2010)]{2010MNRAS.401..160A} Austermann, J.~E., 
et al.\ 2010, \mnras, 401, 160 
\bibitem[B{\'e}thermin et al.(2010)]{2010arXiv1003.0833B} B{\'e}thermin, 
M., Dole, H., Cousin, M., \& Bavouzet, N.\ 2010, arXiv:1003.0833 
\bibitem[Brisbin et al. (2010)] {brisbin2010} Brisbin, D., et al. 2010, MNRAS, submitted
\bibitem[Cantalupo et al.(2010)]{2010ApJS..187..212C} Cantalupo, C.~M., 
Borrill, J.~D., Jaffe, A.~H., Kisner, T.~S., 
\& Stompor, R.\ 2010, \apjs, 187, 212 
\bibitem[Chapin et al.(2010)]{2010arXiv1003.2647C} Chapin, E.~L., et al.\ 
2010, arXiv:1003.2647
\bibitem[Condon(1974)]{1974ApJ...188..279C} Condon, J.~J.\ 1974, \apj, 188, 
279 
\bibitem[Coppin et al.(2005)]{2005MNRAS.357.1022C} Coppin, K., Halpern, M., 
Scott, D., Borys, C., \& Chapman, S.\ 2005, \mnras, 357, 1022
\bibitem[Devlin et al.(2009)]{2009Natur.458..737D} Devlin, M.~J., et al.\ 
2009, \nat, 458, 737 
\bibitem[Dole et 
al.(2006)]{2006A&A...451..417D} Dole, H., et al.\ 2006, \aap, 451, 417 
\bibitem[Downes et al.(1986)]{1986MNRAS.218...31D} Downes, A.~J.~B., 
Peacock, J.~A., Savage, A., \& Carrie, D.~R.\ 1986, \mnras, 218, 31 
\bibitem[Dye et al.(2009)]{2009ApJ...703..285D} Dye, S., et al.\ 2009, 
\apj, 703, 285 
\bibitem[Fadda et al.(2006)]{2006AJ....131.2859F} Fadda, D., et al.\ 2006, 
\aj, 131, 2859
\bibitem[Fernandez-Conde et 
al.(2008)]{2008A&A...481..885F} Fernandez-Conde, N., Lagache, G., Puget, J.-L., \& Dole, H.\ 2008, \aap, 481, 885 
\bibitem[Furusawa et al.(2008)]{2008ApJS..176....1F} Furusawa, H., et al.\ 
2008, \apjs, 176, 1 
\bibitem[Griffin et al.(2010)] {GriffinSPIRE} Griffin, M. et al.,\ 2010, \aap, in press
\bibitem[Hogg(2001)]{2001AJ....121.1207H} Hogg, D.~W.\ 2001, \aj, 121, 1207 
\bibitem[Ivison et al.(2007)]{2007MNRAS.380..199I} Ivison, R.~J., et al.\ 
2007, \mnras, 380, 199 
\bibitem[Lacy et al.(2005)]{2005ApJS..161...41L} Lacy, M., et al.\ 2005, 
\apjs, 161, 41 
\bibitem[Lilly et al.(1999)]{1999ApJ...518..641L} Lilly, S.~J., Eales, 
S.~A., Gear, W.~K.~P., Hammer, F., Le F{\`e}vre, O., Crampton, D., Bond, 
J.~R., \& Dunne, L.\ 1999, \apj, 518, 641
\bibitem[Le Borgne et 
al.(2009)]{2009A&A...504..727L} Le Borgne, D., Elbaz, D., Ocvirk, P., \& Pichon, C.\ 2009, \aap, 504, 727 
\bibitem[Lonsdale et al.(2003)]{2003PASP..115..897L} Lonsdale, C.~J., et 
al.\ 2003, \pasp, 115, 897 
\bibitem[Magnelli et 
al.(2009)]{2009A&A...496...57M} Magnelli, B., Elbaz, D., Chary, R.~R., Dickinson, M., Le Borgne, D., Frayer, D.~T., \& Willmer, C.~N.~A.\ 2009, \aap, 496, 57 
\bibitem[Marsden et al.(2009)]{2009ApJ...707.1729M} Marsden, G., et al.\ 2009, \apj, 707, 1729
\bibitem[Mortier et al.(2005)]{2005MNRAS.363..563M} Mortier, A.~M.~J., et 
al.\ 2005, \mnras, 363, 563 
\bibitem[Nguyen et al. (2010)]{spireconfusion} Nguyen, H. et al., 2010, \aap, in press
\bibitem[Oliver et al.(2010a)] {oliverSurvey} Oliver, S. et al.,\ 2010, in prep
\bibitem[Oliver et al.(2010b)] {oliverCounts} Oliver, S. et al.,\ 2010, \aap, in press
\bibitem[Pascale et al.(2009)]{2009ApJ...707.1740P} Pascale, E., et al.\ 
2009, \apj, 707, 1740 
\bibitem[Patanchon et al.(2008)]{2008ApJ...681..708P} Patanchon, G., et al.\ 2008, \apj, 681, 708 
\bibitem[Patanchon et al.(2009)]{2009ApJ...707.1750P} Patanchon, G., et 
al.\ 2009, \apj, 707, 1750
\bibitem[Pilbratt et al.(2010)]{Herschel} Pilbratt, G., et al.\ 2010, \aap, in press
\bibitem[Pope et al.(2005)]{2005MNRAS.358..149P} Pope, A., Borys, C., Scott, D., Conselice, C., Dickinson, M., \& Mobasher, B.\ 2005, \mnras, 358, 149
\bibitem[Pope et al.(2006)]{2006MNRAS.370.1185P} Pope, A., et al.\ 2006, \mnras, 370, 1185 
\bibitem[Roseboom et al.(2009)]{2009MNRAS.400.1062R} Roseboom, I.~G., Oliver, S., Parkinson, D., \& Vaccari, M.\ 2009, \mnras, 400, 1062 
\bibitem[Savage 
\& Oliver(2007)]{2007ApJ...661.1339S} Savage, R.~S., \& Oliver, S.\ 2007, \apj, 661, 1339 
\bibitem [Scharz(1978)]{1978AS...6..461S} Schwarz, Gideon E.\ 1978, Annals of Statistics 6 (2): 461–464
\bibitem[Scheuer \& Ryle(1957)]{1957PCPS...53..764S} Scheuer, P.~A.~G., \& Ryle, M.\ 1957, Proceedings of the Cambridge Philosophical Society, 53, 764 
\bibitem[Scott et al.(2002)]{2002MNRAS.331..817S} Scott, S.~E., et al.\ 
2002, \mnras, 331, 817
\bibitem[Shupe et al.(2008)]{2008AJ....135.1050S} Shupe, D.~L., et al.\ 
2008, \aj, 135, 1050]
\bibitem[Smith et al.(2010)]{scatpaper} Smith, A.J., et al., in prep
\bibitem[Stark \& Parker(1995)]{} Stark, P.B., \& Parker, R.L., 1995, Comp. Stat. 10., 129-141
\bibitem[Swinyard et al.(2010)] {SPIREcalibration} Swinyard, B., et al.\ 2010, \aap, in press
\bibitem[Tegmark(1997)]{1997ApJ...480L..87T} Tegmark, M.\ 1997, \apjl, 480, 
L87 
\bibitem[Valiante et al.(2010)]{2010arXiv1007.3259V} Valiante, E., et al.\ 
2010, arXiv:1007.3259 
\bibitem[Xu et al.(2001)]{2001ApJ...562..179X} Xu, C., Lonsdale, C.~J., 
Shupe, D.~L., O'Linger, J., \& Masci, F.\ 2001, \apj, 562, 179 
\bibitem[Yun et al.(2008)]{2008MNRAS.389..333Y} Yun, M.~S., et al.\ 2008, 
\mnras, 389, 333 
\end{thebibliography}
\end{document}